\documentclass[a4paper,fleqn,usenatbib]{mnras}

\usepackage{newtxtext,newtxmath}

\usepackage[T1]{fontenc}
\usepackage{ae,aecompl}
\usepackage{pdflscape}

\usepackage{graphicx}	
\usepackage{amsmath}	
\usepackage{amssymb}	
\usepackage{multirow}
\usepackage{gensymb}

\title[X-ray observations of FO Aqr]{X-ray observations of FO Aqr during the 2016 low state}

\author[M. R. Kennedy et al.]{M. R. Kennedy$^{1,2}$\thanks{Contact e-mail: \href{mailto:markkennedy@umail.ucc.ie}{markkennedy@umail.ucc.ie}}, 
P. M. Garnavich$^{2}$,
C. Littlefield$^{2}$,
P. Callanan$^{1}$,
\newauthor
K. Mukai$^{3,4}$,
E. Aadland$^{5}$,
M. M. Kotze$^{6,7}$,
E. J. Kotze$^{6,8}$
\\
$^{1}$Department of Physics, University College Cork, Cork, Ireland \\
$^{2}$Department of Physics, University of Notre Dame, Notre Dame, IN 46556, USA \\
$^{3}$CRESST and X-ray Astrophysics Laboratory, NASA Goddard Space Flight Center, Greenbelt, MD 20771, USA\\
$^{4}$Department of Physics, University of Maryland, Baltimore County, 1000 Hilltop Circle, Baltimore, MD 21250, USA\\
$^{5}$Department of Physics and Astronomy, Minnesota State University, 1104 7th Avenue South, Moorhead, MN 56563, USA  \\
$^{6}$South African Astronomical Observatory, PO Box 9, Observatory 7935, South Africa\\
$^{7}$South African Large Telescope, PO Box 9, Observatory 7935, South Africa\\
$^{8}$Department of Astronomy, University of Cape Town, Private Bag X3, Rondebosch 7701, Cape Town, South Africa
}

\date{Accepted XXX. Received YYY; in original form ZZZ}

\pubyear{2017}

\begin{document}
\label{firstpage}
\pagerange{\pageref{firstpage}--\pageref{lastpage}}
\maketitle

\begin{abstract}
We present the first ever X-ray data taken of an intermediate polar, FO Aqr, when in a low accretion state and during the subsequent recovery. The \textit{Swift} and \textit{Chandra} X-ray data taken during the low accretion state in July 2016 both show a softer spectrum when compared to archival data taken when FO Aqr was in a high state. The X-ray spectrum in the low state showed a significant increase in the ratio of the soft X-ray flux to the hard X-ray flux due to a change in the partial covering fraction of the white dwarf from $>85\%$ to $70^{+5}_{-8}\%$ and a change in the hydrogen column density within the disc from 19$^{+1.2}_{-0.9}\times10^{22}$ cm$^{-2}$ to 1.3$^{+0.6}_{-0.3}\times10^{22}$ cm$^{-2}$. \textit{XMM-Newton} observations of FO Aqr during the subsequent recovery suggest that the system had not yet returned to its typical high state by November 2016, with the hydrogen column density within the disc found to be 15$^{+3.0}_{-2.0}$ cm$^{-2}$. The partial covering fraction varied in the recovery state between $85\%$ and $95\%$. The spin period of the white dwarf in 2014 and 2015 has also been refined to 1254.3342(8) s. Finally, we find an apparent phase difference between the high state X-ray pulse and recovery X-ray pulse of $0.17$, which may be related to a restructuring of the X-ray emitting regions within the system.
\end{abstract}




\section{Introduction}
Intermediate polars (IPs) are a class of cataclysmic variables (CVs) which contain a weakly magnetic white dwarf (WD) primary and a low mass companion. The accretion mechanism varies for individual IP systems. In the marjoity of IP systems, an accretion disc is present up until the magnetic pressure from the WD overcomes the ram pressure in the disc. The material then follows ``accretion curtains'' from the inner edge of the disc to the nearest magnetic pole of the WD \citep{1988MNRAS.231..549R}.

There is evidence that in V2400 Oph, no accretion disc is present (\citealt{1995MNRAS.275.1028B}; \citealt{1997MNRAS.287..117B}; \citealt{2002MNRAS.331..407H}). Instead, material follows a stream from the L1 point and falls ballistically towards the WD until it is caught by the magnetic field of the WD. This is the ``stream-fed'' model \citep{Hameury1986}. Finally, some IPs show evidence of both ``disc-fed'' and ``stream-fed'' accretion simultaneously. This is the ``disc-overflow'' model (\citealt{Lubow1989}; \citealt{Armitage1996}).

Variability in the optical and X-ray light curves of IPs is often used to characterise which accretion mechanism in the system dominates. The spin period of the WD manifests itself as a pulse in the optical and X-ray light curves. In power spectra of IPs, the strongest signals occur at the spin frequency, $\omega$, the orbital frequency $\Omega$, and the beat frequency $\omega-\Omega$ (which can arise due to either accretion via the ``stream-fed'' model or the reprocessing of X-rays by a fixed structure in the systems rest frame). \cite{Wynn1992} and \cite{Ferrario1999} showed how the strength of these peaks can be used to determine the dominant accretion mechanism in the system. For disc accretion (and assuming a perfect up-down symmetry to the magnetic field), the strongest signals are at $\omega$ in both the X-ray and optical power spectra, with additional power at the beat frequency in the optical power spectrum also expected. For stream accretion, the picture is more complicated. The X-ray power spectrum should show strong power at $\omega$, $\omega-\Omega$, $2\omega-\Omega$ and, depending on the inclination of the system, $2\omega$. The optical power spectrum should show dominant power at $\omega$ for high inclination systems ($\sim60 \degree$ or higher) or at $2(\omega-\Omega)$ for lower inclination systems.

Some IPs, such as AO Psc and V1223 Sgr, have been observed to display low states, where the optical magnitude of the IP can decrease by 1-1.5 mag \citep{Garnavich1988}. The cause of these low states is currently suspected to be star spots passing over the L1 point on the surface of the companion star, temporarily halting mass transfer \citep{Livio1994}. This effect has been explored in detail in the system AE Aqr, where it was found that starspots tended to migrate and cluster around the L1 point, causing the observed low states (\citealt{Hill2014}; \citealt{Hill2016}).

FO Aquarii (FO Aqr) is an IP with very strong optical pulsations visible in its optical light curve and an orbital period of 4.8508 hr \citep{Kennedy2016}. Initially discovered as an X-ray source in 1979 \citep{Marshall1979}, it was classified as a cataclysmic variable by \cite{Patterson1983}. It has been dubbed the ``King of the Intermediate Polars'' due to the 0.2 mag amplitude of its optical pulsations \citep{Patterson1983} and, until May 2016, had never been observed in a low state, with observations dating back as far as 1923 \citep{Garnavich1988}. The spin signal of the WD has been seen in both the optical and X-ray light curves of FO Aqr. In the X-rays, the spin period had a value of 1254 s (20.9 mins) in 2003 \citep{Evans2004}. The value of the spin period observed in the optical has had a complex history (for a full history, see \citealt{Kennedy2016} and references within). The most recent value for the spin period was 1254.3401 s \citep{Kennedy2016}. The beat period was also detected at 1354.329 s, and the WD was proposed to be spinning down.

Previous X-ray observations of FO Aqr using the \textit{XMM-Newton} space telescope revealed a complex X-ray spectrum. \cite{Evans2004} observed FO Aqr during its typical high state in 2001, and found a multi-temperature plasma model combined with partial absorbers that had covering fractions as high as 0.94 could describe the spectrum well. They also noted that the ratio of soft to hard X-rays varied over the 4.8508 hour orbital period, suggesting material within the accretion disc or the stream-disc impact region periodically obscured the WD.  \cite{2005A&A...439..213P} performed a study of IPs, and found an orbital modulation in the X-rays is common among IPs. The cause of the orbital modulation is thought to be due to photoelectric absorption by material in the impact region between the ballistic stream and the accretion disc. \cite{2012AJ....144...53P} also noted this behaviour in FO Aqr, and showed how the X-ray spectrum varied over the orbital period.

In April of 2016, FO Aqr was found to be at a V-band magnitude of 15.7 \citep{Littlefield2016a}, compared to its normal V-band magnitude of 13.6. Detailed analysis of the optical photometry taken in this low state revealed that the strongest signal in the power spectrum of FO Aqr was no longer the spin period of 20.9 min, but rather both the spin signal and a signal with a period of 11.26 mins, which is half the beat period of the system \citep{littlefield2016b}, had equal power. Analysis of this power spectrum led \cite{littlefield2016c} to conclude that the accretion geometry in FO Aqr had transitioned from a disc-fed geometry to a ``disc-overflow'' geometry, where the disc-fed accretion was generating the spin signal while the stream-fed accretion was generating the half beat signal. This is not the first time that the ``disc-overflow'' geometry has been applied to FO Aqr (\citealt{Hellier1993}; \citealt{1996MNRAS.280..937N}). Evidence for ``disc-overflow'' accretion has also been seen in previous X-ray observations of FO Aqr and the X-ray pulse has been seen to change shape depending on the accretion geometry \citep{1998MNRAS.297..337B}.

The photometry presented in \cite{littlefield2016c} also revealed that, during the low state, the dominant power in the power spectrum was orbital dependant, with a 22.54 minute period dominating from orbital phase $0.9<\phi<1.4$ and the 11.26 minute period dominating from $0.4<\phi<0.9$. Phasing the light curves on the known orbital period of the system also revealed the presence of the eclipse around orbital phase 0. This eclipse has been seen previously in FO Aqr during its high state, and is thought to be a grazing eclipse of the outer accretion disc \citep{Hellier1989}. The appearance of this eclipse in the photometry presented by \cite{littlefield2016c} suggests the accretion disc was still present during the low state. 

Here, we present X-ray observations of FO Aqr in its first confirmed low state and during its subsequent recovery.

\section{Observations}
\begin{table*}
	\centering
	\caption{Details of the various X-ray observations of FO Aqr in 2016. The average count rate of the X-ray observations from \textit{Swift} and \textit{XMM-Newton} are in the 0.3-10 keV band and the \textit{Chandra} count rate is in the 0.3-7 keV band.}
	\begin{tabular}{l c c c c}
		\hline
        Date            & Telescope                         & Average Flux              & Duration                                         & Phase Coverage\\
		(YYYY-MM-DD)	&                                   &                           &                                                  & \\
		\hline
		X-ray           &                                   & counts s$^{-1}$           & ks                                               & \\
		\hline\hline
		$2016-07-10$	& \textit{Swift}                             & 0.4$\pm$3                 & 1.545                                            &0.11-0.15\\
		$2016-07-11$	& \textit{Swift}                             & 0.27$\pm$0.08             & 1.19                                             &0.77-0.80\\
		$2016-07-12$	& \textit{Swift}                             & 0.24$\pm$0.12             & 1.5                                              &0.06-0.09\\
		$2016-07-13$	& \textit{Swift}                             & 0.25$\pm$0.08             & 0.075                                            &0.61-0.61\\
		$2016-07-14$	& \textit{Swift}                             & 0.28$\pm$0.10             & 0.73                                             &0.95-0.99\\
		$2016-07-16$	& \textit{Swift}                             & 0.46$\pm$0.13             & 1.48                                             &0.28-0.32\\
		$2016-07-26$	& \textit{Chandra}                           & 1.3$\pm$0.9               & 18.02                                            &0.00-1.00\\
		$2016-11-13$    & \textit{XMM-Newton}                        & 2.6$\pm$1.6               & 45.6                                             &0.00-1.00\\
		\hline
	\end{tabular}
	\label{tab:observations}
\end{table*}

\subsection{\textit{Swift}}
Target of Opportunity (ToO) observations of FO Aqr were carried out using the \textit{Swift} Gamma-Ray Burst Telescope for a total of 6.5 ks over several observations as detailed in Table~\ref{tab:observations}. The observations were taken using the X-ray Telescope (XRT; \citealt{2000SPIE.4140...64B}; \citealt{2000SPIE.4140...87H}) instrument operated in photon counting (PC) mode. The data were obtained from the \textit{Swift}-XRT data products generator\footnote{\url{http://www.swift.ac.uk/user_objects/}}. The average flux from FO Aqr in the 0.3-10 keV energy range was 0.3-0.4 counts s$^{-1}$ for the 2006 observations during the high state. The low state had a similar count rate in the \textit{Swift} data of 0.2-0.4 counts s$^{-1}$. The average V-band magnitude of FO Aqr during these observations was 14.9$\pm$0.2.

\subsection{\textit{Chandra}}
X-ray observations of FO Aqr were carried out for a total of 18.02 ks starting 2016 Jul 26 7:26AM (UT) (ObsID \#18889) using the ACIS-S instrument \citep{2003SPIE.4851...28G}. To mitigate pileup, ACIS-S  was run in continuous clocking (CC) mode. CC mode allows for higher time resolution for observations, at the cost of a spatial dimension in the data (rather than the typical 1024$\times$1024 pixel image from a \textit{Chandra} observation, the resultant image from the CC-mode observation was 1024$\times$1 pixel, since one of the spatial dimensions had been collapsed). Source and background regions were extracted from the 1024$\times$1 pixel image. Data reprocessing was carried out using CIAO 4.8.1 \citep{2006SPIE.6270E..1VF} and CALDB 4.7.2. Again, the average V-band magnitude of FO Aqr during these observations was 14.9$\pm$0.2.

\subsection{XMM Newton}
FO Aqr was observed by \textit{XMM-Newton} on 2016 Nov 13 20:00 (UT) for a total of 45.6 ks. The EPIC-pn \citep{2001A&A...365L..18S} and EPIC-MOS \citep{2001A&A...365L..27T} instruments were operated in Small Window mode with medium filters inserted. The RGS spectrographs (\citealt{1998sxmm.confE...2B}; \citealt{2001A&A...365L...7D}) were operated in their normal spectroscopy modes. The optical monitor (OM; \citealt{2001A&A...365L..36M}) was operated in fast timing mode with the UVW1 filter (effective wavelength$=291$ nm, width$=83$nm) inserted for the first half of the observations, and the UVM2 filter (effective wavelength$=231$ nm, width$=48$nm) inserted for the second half of the observation. The data were reduced and analysed using the \textit{XMM-Newton} {\sc Science Analysis Software} ({\sc SAS} v15.0.0; \citealt{XMMHandbook}). The data from the EPIC instruments were processed using the {\sc emproc} and {\sc epproc} tasks. The files used for generating light curves were corrected to Barycentric Julian Date (BJD) using the command {\sc Barycen}. All events were used for generating light curves, while only events which were recorded during good time intervals (GTI) were used for generating spectra. The average V-band magnitude of FO Aqr during the 2016 $XMM-Newton$ observations was 13.7$\pm$0.2, which is 0.2 magnitudes fainter than FO Aqr would have been during the archive 2001 \textit{XMM-Newton} observations \citep{Evans2004} and a magnitude brighter than during the \textit{Chandra} observations.

\subsubsection{Ground-based Photometry}
Simultaneous ground-based optical photometry observations of FO Aqr during the 2016 \textit{XMM-Newton} observations were also obtained. FO Aqr was observed in the V-band using the Sutherland High-Speed Optical Camera (SHOC) on the South African Astronomical Observatory (SAAO) 1m telescope. A total of 11355 images were taken with a cadence of 1s between 18:15 UTC and 21:25 UTC with the Johnson V-band filter inserted. FO Aqr was also observed using the 80 cm Sarah L. Krizmanich Telescope (SLKT) at the University of Notre Dame. 2724 images were taken with no filter inserted, with a typical cadence of 6s. Finally, a total of 1927 images of FO Aqr were taken with a mean cadence of 21s by various members of the American Association of Variable Star Observers (AAVSO) during the \textit{XMM-Newton} observations. 704 of these images were taken in the Johnson V-band, and the rest were taken without a filter inserted.

\section{Analysis}

\subsection{Timing Analysis}

\subsubsection{The Spin Period of FO Aqr}



The spin ephemeris used by \cite{littlefield2016c} is defined such that maximum optical light from the spin pulse should occur at spin phase 1.0. However, when this ephemeris was used to phase the optical data presented here, the phase of maximum light in the spin-phased optical light curve occured at phase 0.75. \cite{littlefield2016c} encountered this same issue, and tentatively suggested that the arrival phase of maximum light seemed to be related to the overall brightness of the system (Fig 5 of their paper). Based on their proposal, we would expect the phase of maximum light to be around spin phase 0.95, given FO Aqr's optical magnitude of 13.7 during the \textit{XMM-Newton} observations. Since the predicted and observed phases do not agree, an alternate explanation is necessary.

Our re-analysis suggests that the spin period in the ephemeris used by \cite{littlefield2016c}, which was taken from an analysis of the \textit{K2} photometry by \citet{Kennedy2016}, was too large. As Kennedy et al. pointed out, the pulse arrival times during the high state were closely correlated with the system's brightness; thus, when the system brightened for the final third of the \textit{K2} observations, the spin pulse moved to a later phase. To measure how this impacts the measured spin period, we split the \textit{K2} data into two bins based on FO Aqr's brightness, and the measured spin period in each bin (1254.3355 sec) is shorter than the global period (1254.3401(4) sec), implying that the phase shift of the spin pulse introduced a small error into the measured period.

As an additional test of this result, we simulated the \textit{K2} light curve with a piecewise light curve consisting of four consecutive, non-overlapping sinusoids. All four sinusoids had the same true period (1254.3355 sec), but the second and fourth segments had a phase offset of 0.02 with respect to the first and third segments. The length of each segment and its phase shift were approximated from the \textit{K2} data, and we matched the sampling rate to the median \textit{K2} sampling rate. Power spectra of the simulated bright data (the second and fourth segments) and of the simulated faint data (the first and third segments) both successfully recovered the true input period. However, a power spectrum of the full simulated dataset showed an inflated period of 1254.3407 sec. Thus, our simulations validate our re-analysis of the \textit{K2} spin period, demonstrating that even a relatively small phase shift can interfere with an accurate determination of the spin period. Simple power spectral analysis of a lengthy dataset is not a reliable method of measuring FO Aqr's true spin period; the brightness of the system must be taken into account.

\begin{figure}
	\includegraphics[width=\columnwidth]{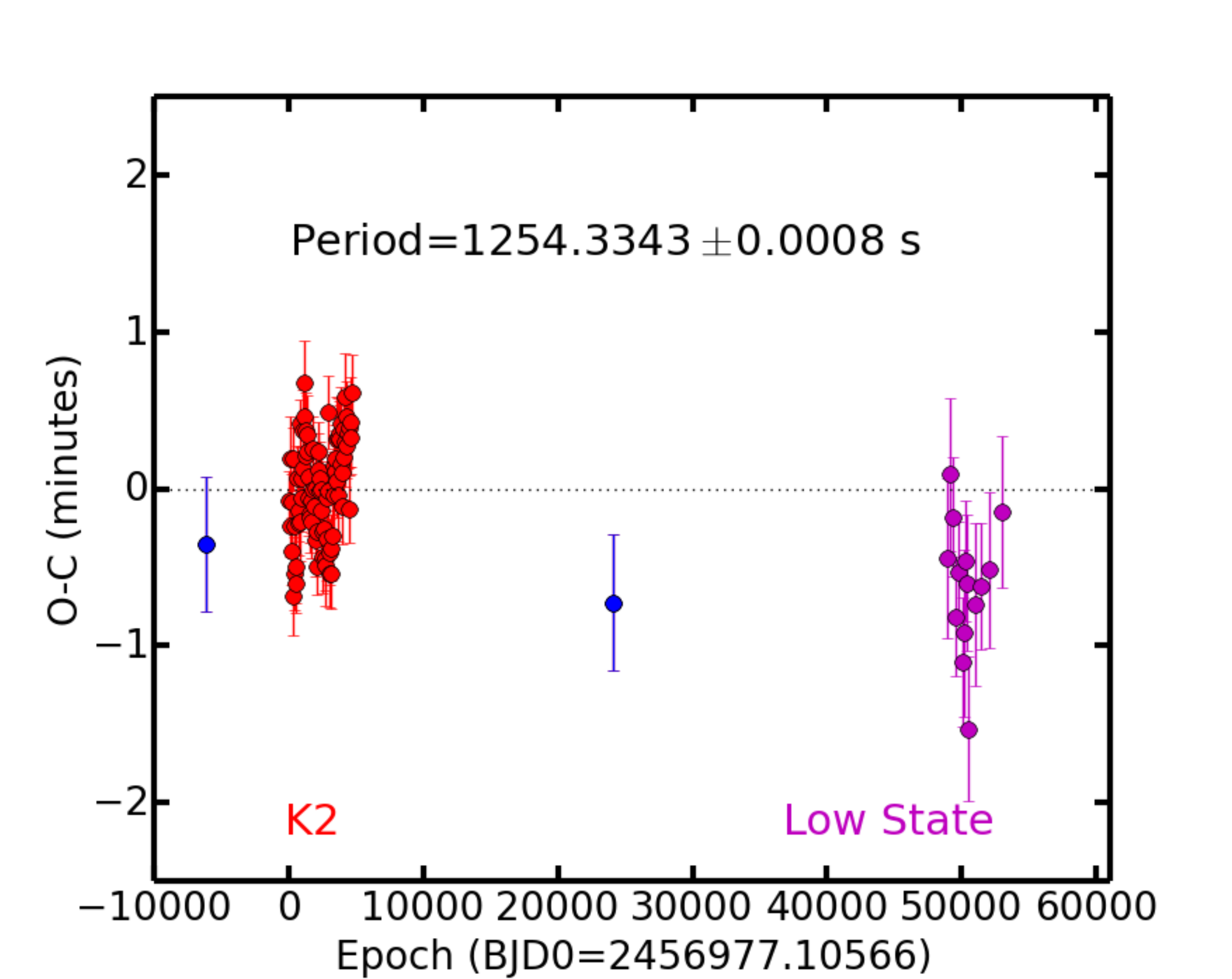}
    \caption{The residuals to the new ephemeris. The \textit{K2} measurements (red) show a coherent variation that correlates with brightness changes. The blue points are Bonnardeau (2016) measurements for 2014 and 2015. The timing measurements at the
    end of the low state were not used to calculate the new ephemeris, but are clearly consistent with an extrapolation into 2016.}
    \label{fig:ephem}
\end{figure}

 We have calculated a new ephemeris that combines the \textit{K2} data with spin timings from \cite{2016IBVS.6181....1B}. We divide up the \textit{K2} data into 85 sections each containing 50 spin cycles and fit a linear ephemeris to each group. The phase shift that correlates with the brightness variations is easily seen in this binned data and results in a root-mean-squared (RMS) variation of 0.25 minutes over the 85 independent measurements. This can be considered the uncertainty floor on any spin timing unless there is a correction for the brightness/phase correlation. We increase the 2014 and 2015 \cite{2016IBVS.6181....1B} error estimates to 0.25 minutes and find a new ephemeris of

\begin{equation} \label{eqn:SpinEph}
    T(BJD) = 2456977.1057(8)+E*0.014517757(9)
\end{equation}

which is a period of 1254.3342(8) seconds. This is consistent with the \textit{K2} spin period found by splitting the measurements by brightness. We show in Fig.~\ref{fig:ephem} the residuals to the ephemeris. We include spin timing measurements obtained toward the end of the faint state, when individual spin pulses were sufficiently strong to be easily identifiable in optical light curves. These spin timings are consistent with this new ephemeris and suggests no major phase shift occurred across the low state. Taken in isolation, the spin timing residuals shown in Fig.~\ref{fig:ephem} imply an even shorter period would be a better fit to the 2014 to 2016 data. However, the long-term period changes seen in FO~Aqr require high-order polynomial solutions that are beyond the scope of the current data.

When propagated across the two-year span between the \textit{K2} observations and the low state, the error accumulates into a large phase shift ($\sim$0.2 in phase) of the spin pulse towards earlier phases, similar to the one observed in \cite{littlefield2016c}. There might have been a true phase shift during the low state, but the one reported in that paper was not physical in origin.



\subsubsection{X-Ray}
\begin{figure}
	\includegraphics[width=\columnwidth]{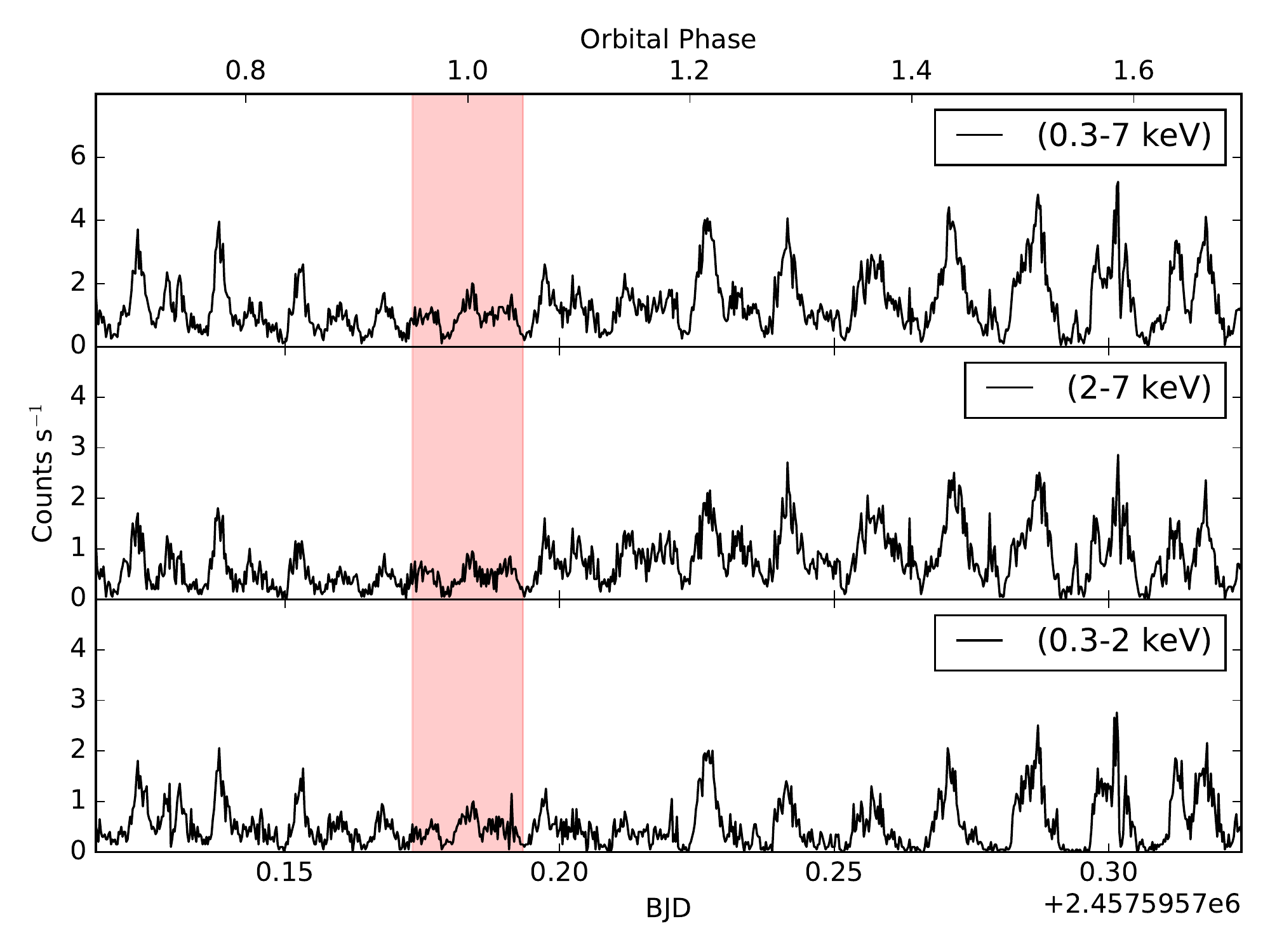}
    \caption{The full, hard (2-7 keV) and soft (0.3-2 keV) light curves taken by \textit{Chandra} of FO Aqr in the low state in July 2016. The count rate in the soft and hard X-rays is comparable, and both components contribute nearly equally to the variability. This behaviour is very different to the X-ray light curves taken by \textit{XMM-Newton} in 2001, when FO Aqr was in its high state. The highlighted region shows the data taken between orbital phase 0.95 and 0.05, which is when the grazing eclipse of the system has been observed previously. The orbital phase was calculated as described in the text.}
    \label{fig:chandra_lc}
\end{figure}

\begin{figure}
	\includegraphics[width=\columnwidth]{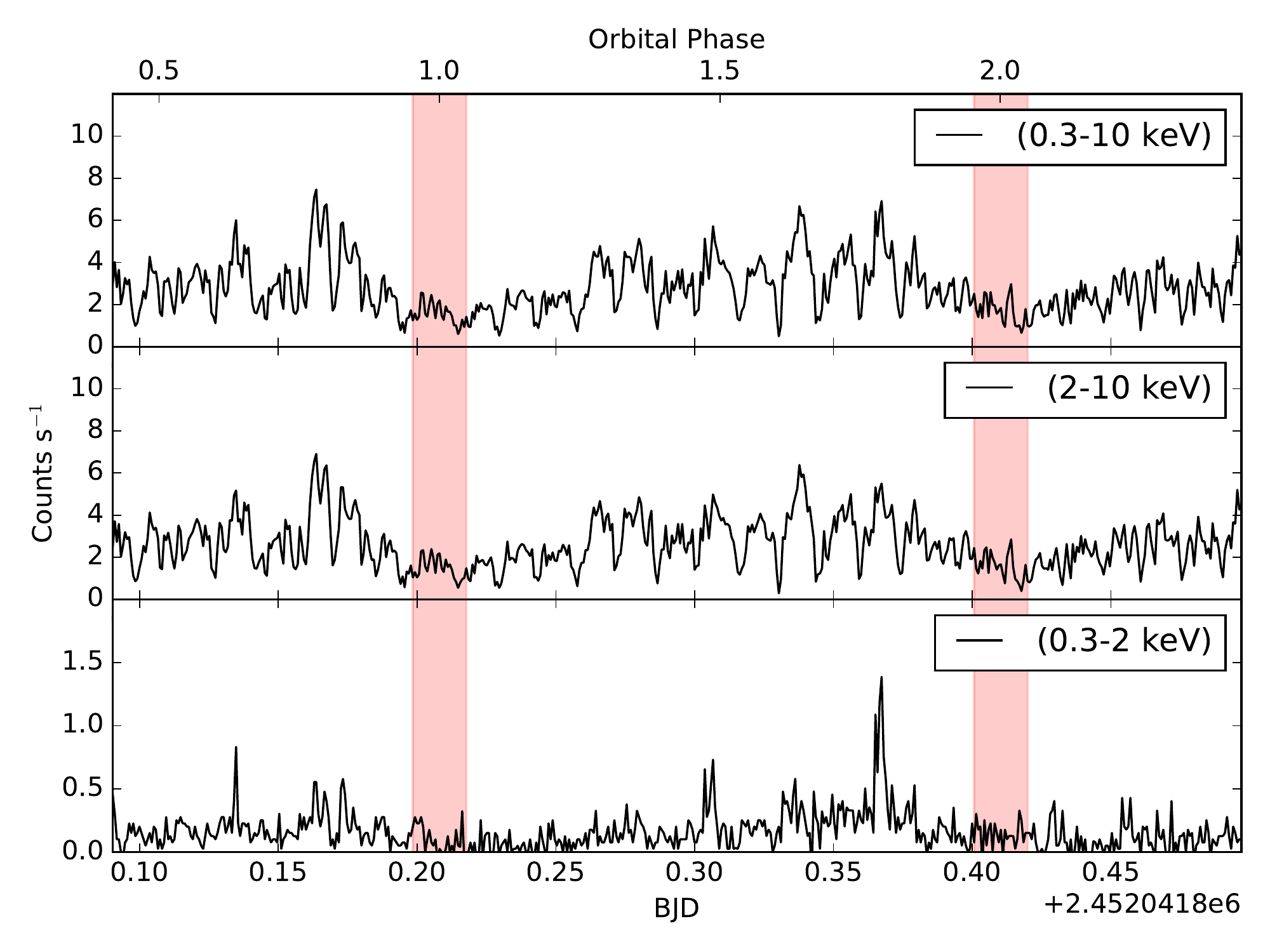}
    \caption{The full, hard (2-10 keV) and soft (0.3-2 keV) light curves from 2001, taken using the EPIC-pn instrument onboard \textit{XMM-Newton} when the system was in the high state. The count rate in the soft X-rays is much lower than in the hard, and the variability of the source is only weakly detected in the soft X-rays.}
    \label{fig:xmm_lc_01}
\end{figure}

\begin{figure}
	\includegraphics[width=\columnwidth]{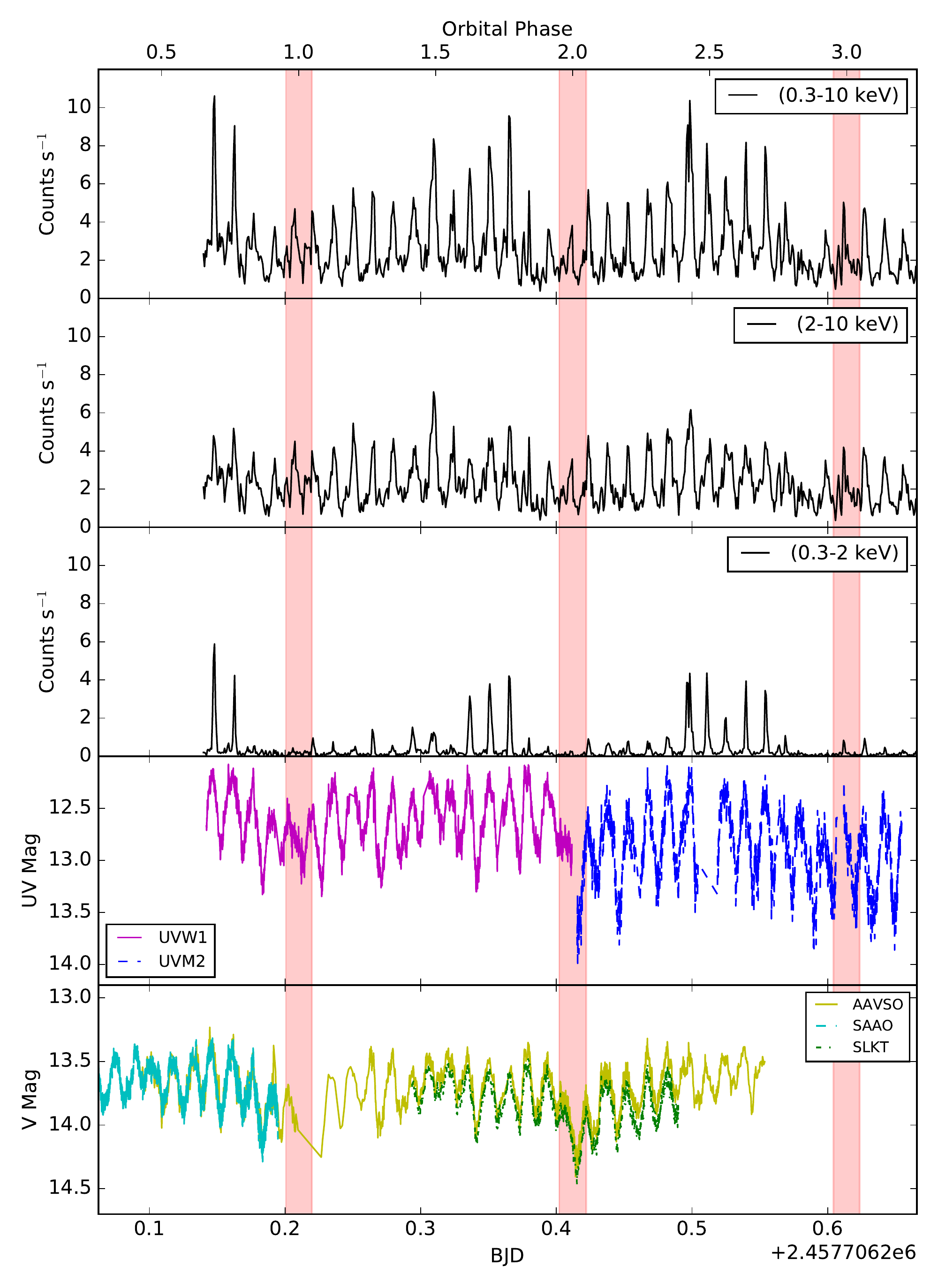}
    \caption{The full (1st panel), hard (2-10 keV; 2nd panel) and soft (0.3-2 keV; 3rd panel) X-ray light curves from 2016, taken using the EPIC-pn instrument onboard \textit{XMM-Newton} when the system was returning to the high state along with the UV light curve from the OM (4th panel), the ground based V-band data taken by AAVSO observers (5th panel, yellow), the SAA0 1m telescope (5th panel, magenta) and the SLKT (5th panel, green). The variability in the soft X-rays lies somewhere between the high variability seen in the low state \textit{Chandra} data and the near-constant flux seen in the 2001 \textit{XMM-Newton} data. The optical and UV pulses arrived at the same phase as the X-ray pulses. The highlighted region again shows data taken between orbital phase 0.95 and 0.05.}
    \label{fig:xmm_lc_16}
\end{figure}

The X-ray light curves of FO Aqr taken by \textit{Chandra} in 2016 (Fig.~\ref{fig:chandra_lc}), \textit{XMM-Newton} in 2001 (Fig.~\ref{fig:xmm_lc_01}) and \textit{XMM-Newton} in 2016 (Fig.~\ref{fig:xmm_lc_16}) show very different behaviours. An X-ray pulse was visible in the full, soft (0.3-2 keV) and hard (2-7 keV) light curves from the \textit{Chandra} observations of FO Aqr when the system was in the low state, with most hard X-ray pulses having a soft X-ray counterpart. The archived \textit{XMM-Newton} data from 2001 shows a very different behaviour, with the X-ray pulses only visible in the hard X-rays, with little obvious variability in the soft X-rays. The 2016 \textit{XMM-Newton} light curves, taken when the system was returning to the high state (which for the rest of this paper is referred to as the ``recovery'' state), showed variability somewhere in between the low and high states, with some hard X-ray pulses having a soft X-ray counterpart, while other hard X-ray pulses had no detectable counterparts. The orbital phases of the \textit{Chandra} and 2016 \textit{XMM-Newton} data were calculated using the ephemeris of

\begin{equation}
    T(HJD) = 2456982.2278 + 0.20205976(E)
\end{equation}

which is a combination of the mid-eclipse time taken from \citep{Kennedy2016} and the orbital period from \citep{Marsh1996}, while the orbital phases of the 2001 \textit{XMM-Newton} data were calculated using the ephemeris of

\begin{equation}
    T(HJD) = 2452041.806 + 0.20205976(E)
\end{equation}

which is a combination of the time of minimum light taken from \citep{Evans2004} and the orbital period from \citep{Marsh1996}.

\begin{figure*}
	\includegraphics[width=\textwidth]{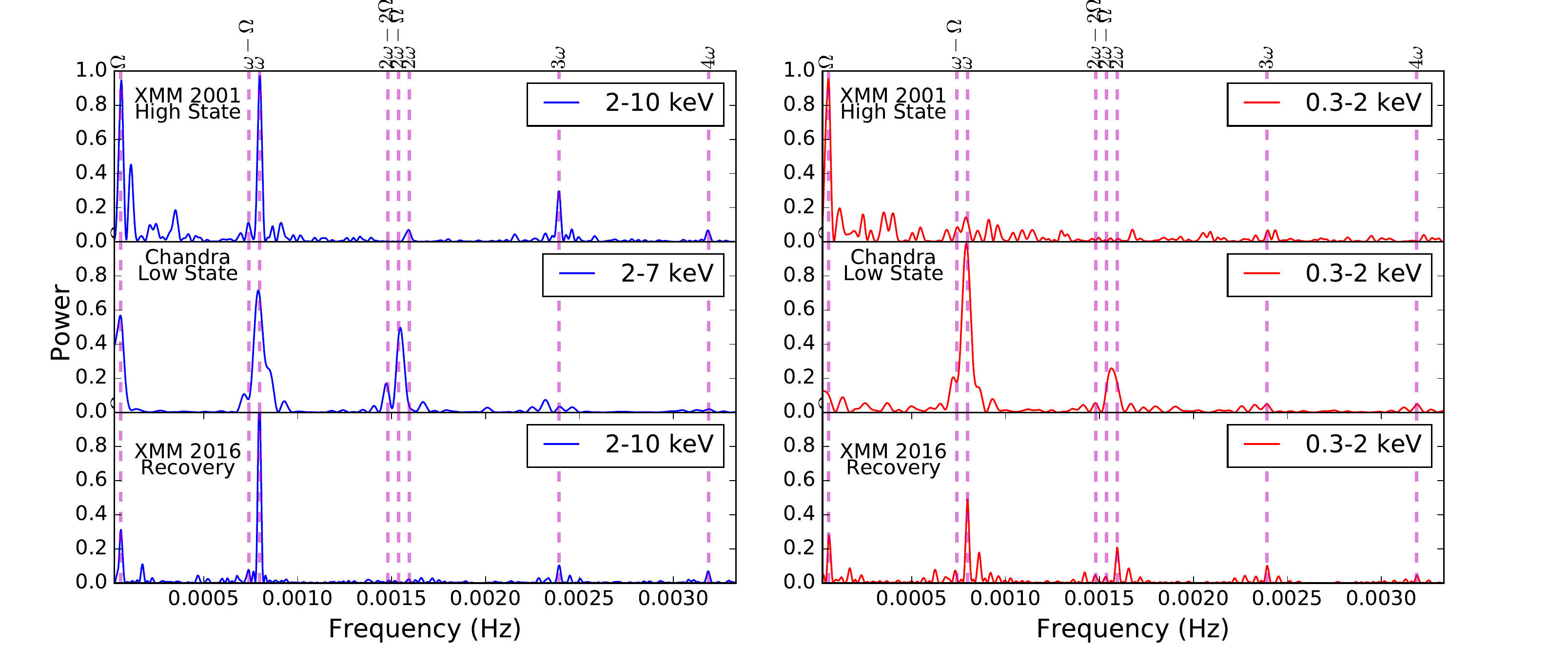}
    \caption{A Lomb Scargle periodogram (LSP) of the archive \textit{XMM-Newton} (top), \textit{Chandra} (middle) and 2016 \textit{XMM-Newton} (bottom) light curves. The spin frequency ($\omega$) shows significant power in all three data sets in the 2-10 keV band. However, hard (left plots) and soft (right plots) X-ray contributions to the various peaks in the LSP vary significantly between the high, low and recovery states. To make direct comparison easier between the three data sets, the power of each spectrum has been normalised such that the strongest peak in power spectrum of the combined soft and hard data had a power of 1.}
    \label{fig:xray_power}
\end{figure*}

To understand how the variability in the soft and hard X-rays has changed, a Lomb-Scargle periodiogram (LSP; \citealt{Lomb76}; \citealt{scargle82}) was applied to the light curves to search for the strongest frequencies in each energy band and observation set. The LSP of the \textit{Chandra} and XMM observations is shown in Fig.~\ref{fig:xray_power}, and highlights the dramatic change in the variability of the soft X-rays coming from FO Aqr.

\subsection{Spectral Analysis}

\subsubsection{Low State}
\begin{figure}
	\includegraphics[width=\columnwidth]{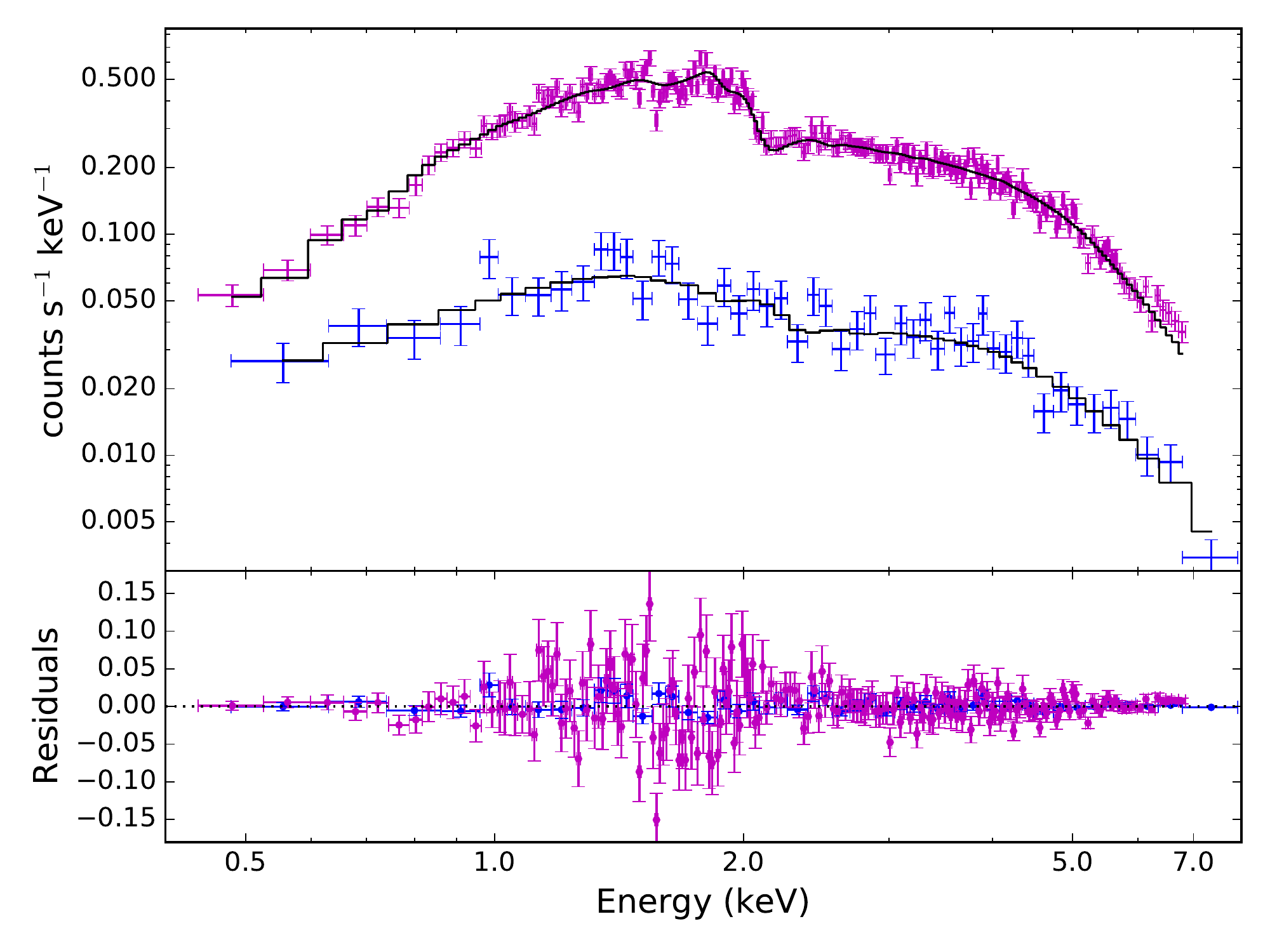}
    \caption{The observed X-ray spectrum of FO Aqr during its low state taken with the ACIS instrument onboard the \textit{Chandra} spacecraft (magenta, top) and using the XRT instrument on \textit{Swift} (blue, bottom). The black histogram shows the best fit models as described in Table~\ref{tab:model_data}.}
    \label{fig:model_chandra}
\end{figure}

The full X-ray spectrum from 0.3-7.0 keV was extracted from the \textit{Chandra} ACIS-CC data using the CIAO analysis software, and analysed using the HEASARC software package {\sc Xspec} (version 12.9.0). Fig.~\ref{fig:model_chandra} shows the X-ray spectrum from the full exposure. 

\cite{Evans2004} previously fit the \textit{XMM-Newton} PN and MOS X-ray spectrum using the complex model of an interstellar absorber,  two partial absorbers (one representing absorption by the accretion curtains and one representing absorption by the accretion disc) and three single temperature thin thermal plasma ({\sc Mekal}) components, alongside various {\sc Gaussian} components to model emission lines. The resulting fit, shown in Fig. 3 and described in Table 1 of \cite{Evans2004}, models the spectrum very well, with a $\chi^{2}_{R}=1.03$.

Initially, the complex model of \cite{Evans2004} was fit to the low state X-ray spectrum ({\sc Xswabs*Xspcfabs*Xspcfabs*(Xsmekal+Xsmekal+\\Xsmekal)} in {\sc Xspec}). While this model fit the spectrum well, with a $\chi^{2}_{R}=0.788$, many of the fit parameters were very poorly constrained, and significantly different to \cite{Evans2004}. Most notable were the lower absorption columns in both of the partial absorber components, alongside the unbounded covering fractions, and also the lack of a constraint on the highest temperature {\sc Mekal} component. 

The \textit{Chandra} spectrum was next fit using a very basic model common to many IPs: a simple interstellar absorber ({\sc Tbabs}), partial absorber ({\sc PartCov*Tbabs}) and a single temperature thin thermal plasma ({\sc Mekal}). The resulting fit, given in Table.~\ref{tab:model_data} and shown in Fig.~\ref{fig:model_chandra}, described the low state X-ray spectrum well. 

\begin{table*}
	\centering
	\caption{The best fit models to the 2016 \textit{Chandra}, \textit{Swift} and \textit{XMM-Newton} data, and the 2006 \textit{Swift} data. Errors quoted are at the 3$\sigma$ level. The plasma temperature for the \textit{Swift} 2006 modelling of FO Aqr was fixed at 29.7 keV based on IBIS observations \citep{Landi2009}.}
	\begin{tabular}{r  c c c c c}
		\hline
                                        &                                   & \textit{Chandra}                                           & \multicolumn{2}{c}{\textit{Swift}}                                         & XMM\\
		Component 						& Parameter							& 2016   	                                        & 2016 		                            & 2006                      & 2016\\
		\hline\hline
		{\sc tbabs (interstellar)}		& $N_{H}$ (10$^{22}$ cm$^{-2}$)		& 0.02$^{+0.08}_{-0.02}$	                        & 0.05$^{+0.2}_{-0.05}$	                & 0.78$^{+0.19}_{0.18}$     & 0.09$^{+0.06}_{-0.04}$\\
		\\
		{\sc tbabs (circumstellar 1)}	& $N_{H}$ (10$^{22}$ cm$^{-2}$)		& 1.3$^{+0.6}_{-0.3}$		                        & 1.5$^{+2.7}_{-0.7}$	                & 7.2$^{+1.2}_{0.9}$        & 15$^{+3.0}_{-2.0}$\\ 
		\\
		{\sc partcov 1}					& cvf								& 0.70$^{+0.05}_{-0.08}$	                        & 0.70$^{+0.15}_{-0.26}$		        & 0.82$^{+0.04}_{-0.03}$    & 0.71$^{+0.04}_{-0.05}$\\
		\\
		{\sc tbabs (circumstellar 2)}	& $N_{H}$ (10$^{22}$ cm$^{-2}$)		& -		                                            & -                 	                & -                         & 3.6$^{+0.6}_{-0.5}$\\ 
		\\
		{\sc partcov 2}					& cvf								& -	                                                & -		                                & -                         & 0.89$\pm$0.02\\
		\\
		\multirow{3}{*}{\sc Mekal}		& kT (keV)							& $>$15			                                    & $>$7		                            & -                         & $>$28\\
		                                & Abundance                         & 1.0                                               & 1.0                                   & -                         & 0.5$^{+0.3}_{-0.2}$\\
										& norm		        	            & 0.160$^{+0.365}_{-0.094}$	                        & 0.83$^{+1.5}_{-?}$			        & -                         & 0.037$^{+0.002}_{-0.002}$\\
		\\
		\multirow{3}{*}{\sc Mekal}		& kT (keV)							& -			                                        & -		                                & -                         & 0.11$^{+0.03}_{-0.02}$\\
		                                & Abundance                         & -                                                 & -                                     & -                         & 0.5$^{+0.3}_{-0.2}$\\
										& norm		        	            & -	                                                & -			                            & -                         & 0.025$^{+0.05}_{-0.02}$\\
		{\sc Bbody}                     & kT (keV)                          & -                                                 & -                                     & 61$^{+8}_{-6}$            & - \\
		{\sc Bremss}                    & kt (keV)                          & -                                                 & -                                     & [29.7]                    & - \\
		\\
		\multirow{3}{*}{\sc Gaussian}	& Centre							& -			                                        & -		                                & -                         & 6.49$^{+0.04}_{-0.02}$\\
		                                & Sigma                             & -                                                 & -                                     & -                         & 0.17$^{+0.04}_{-0.05}$\\
										& norm ($\times10^{-4}$)            & -	                                                & -			                            & -                         & 1.3$^{+0.3}_{-0.3}$\\
		
										& $\chi^{2}_{R}$					& 1.14		                                        & 1.01			                        & 0.92                      & 1.36 \\
		\hline
	\end{tabular}
	\label{tab:model_data}
\end{table*}

\begin{figure}
	\includegraphics[width=\columnwidth]{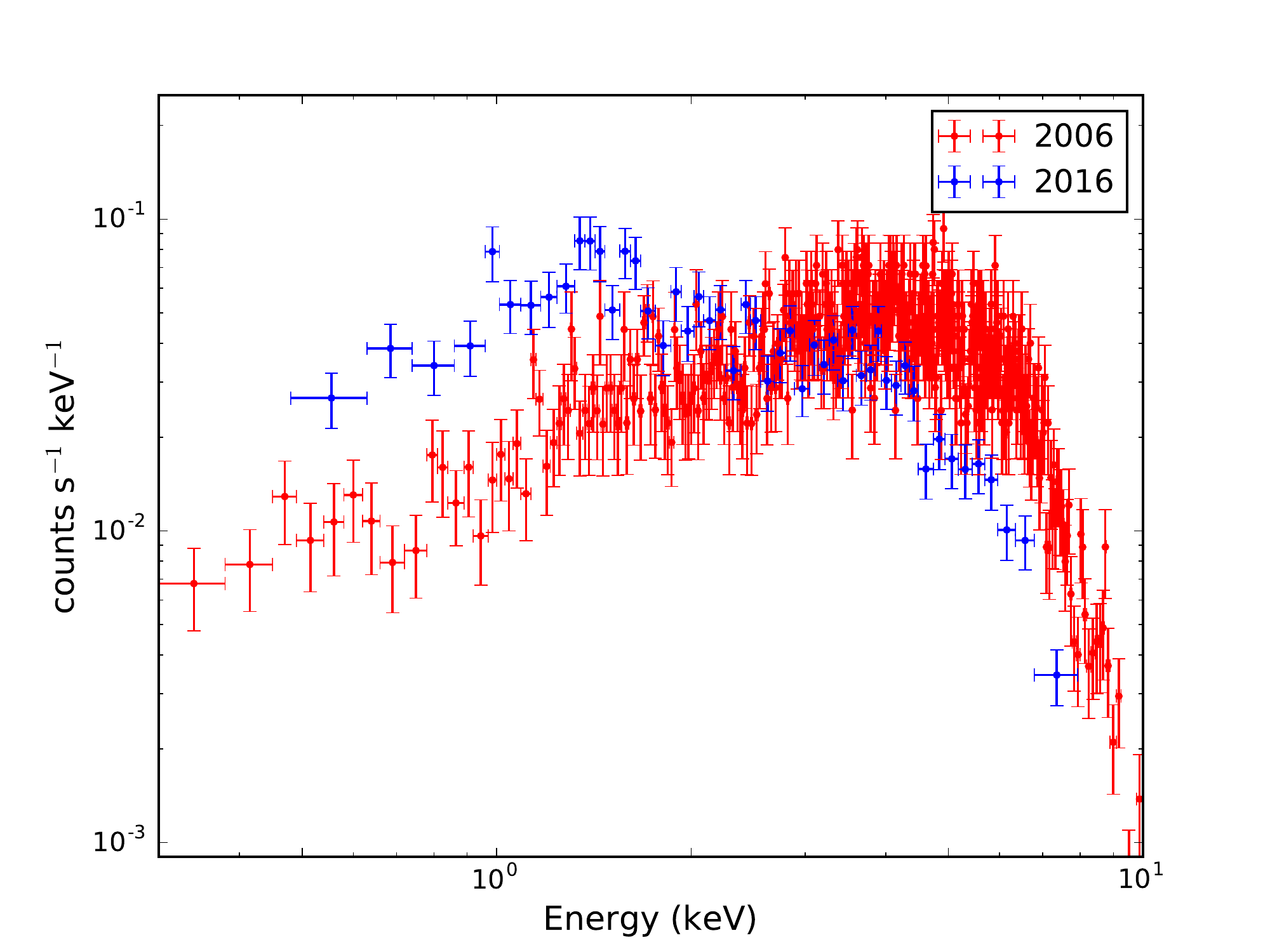}
    \caption{The average X-ray spectrum of FO Aqr taken using the \textit{Swift} XRT during the 2 different states. The 2006 data (red) were taken during the high state, while the 2016 data were taken during the low state.}
    \label{fig:comb_swift_spec}
\end{figure}

Fig.~\ref{fig:comb_swift_spec} shows the average spectrum from the 2006 and 2016 observations of FO Aqr using the XRT instrument on the \textit{Swift} spacecraft. Despite the similar X-ray count rates detected by \textit{Swift} between 2006 and 2016, it is immediately obvious that the energy distribution has changed, with the low state spectrum appearing softer than the high state.

The 2016 spectrum was modelled using a similar model to that used for the \textit{Chandra} data, and the resulting best fit is shown in Fig.~\ref{fig:model_chandra}, with the best fit parameters listed in Table~\ref{tab:model_data}. The parameters from both the \textit{Chandra} and \textit{Swift} modelling are consistent with each other. The unabsorbed flux in the 0.3-10 keV band was estimated from the models in {\sc Xspec} to be $(3.1\pm0.2)\times10^{-11}$ erg cm$^{-2}$ s$^{-1}$ for the \textit{Chandra} data and $(2.6\pm0.4)\times10^{-11}$ erg cm$^{-2}$ s$^{-1}$ for the \textit{Swift} data.

The high state spectrum of FO Aqr taken by \textit{Swift} in 2006 was previously modelled using an absorbed, partially covered Bremsstrahlung and black body model \citep{Landi2009}. The best fit parameters from this model are listed in Table~\ref{tab:model_data}.

\subsubsection{Recovery State} \label{sec:XMM_Model}
\begin{figure}
	\includegraphics[width=\columnwidth]{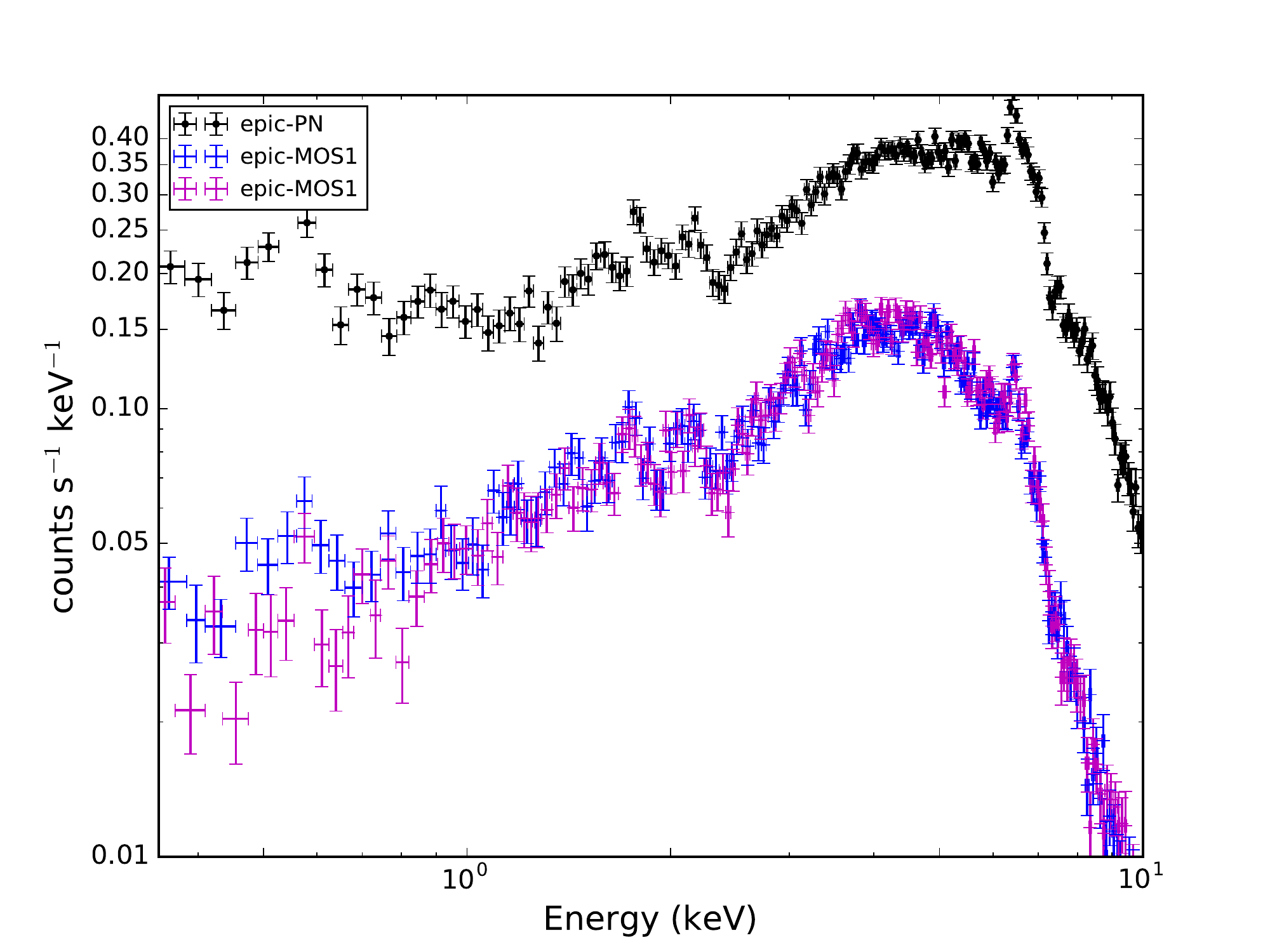}
    \caption{The observed X-ray spectrum of FO Aqr during its recovery state taken with the EPIC-pn, -MOS1 and -MOS2 instruments onboard the \textit{XMM-Newton} spacecraft.}
    \label{fig:raw_xmm}
\end{figure}

The full X-ray spectrum from 0.3-10.0 keV was extracted from the \textit{XMM-Newton} EPIC-pn, -MOS1 and MOS2 instruments and analysed using the HEASARC software package {\sc Xspec} (version 12.9.0). Fig.~\ref{fig:raw_xmm} shows the X-ray spectrum extracted using events recorded during good time intervals only.

As with the low state data, the recovery spectrum was initially fit using the model of \cite{Evans2004}. The spectra from the EPIC-pn, -MOS1 and -MOS2 spectra were fit simultaneously, allowing for a constant calibration factor between the instruments.

\begin{figure}
	\includegraphics[width=\columnwidth]{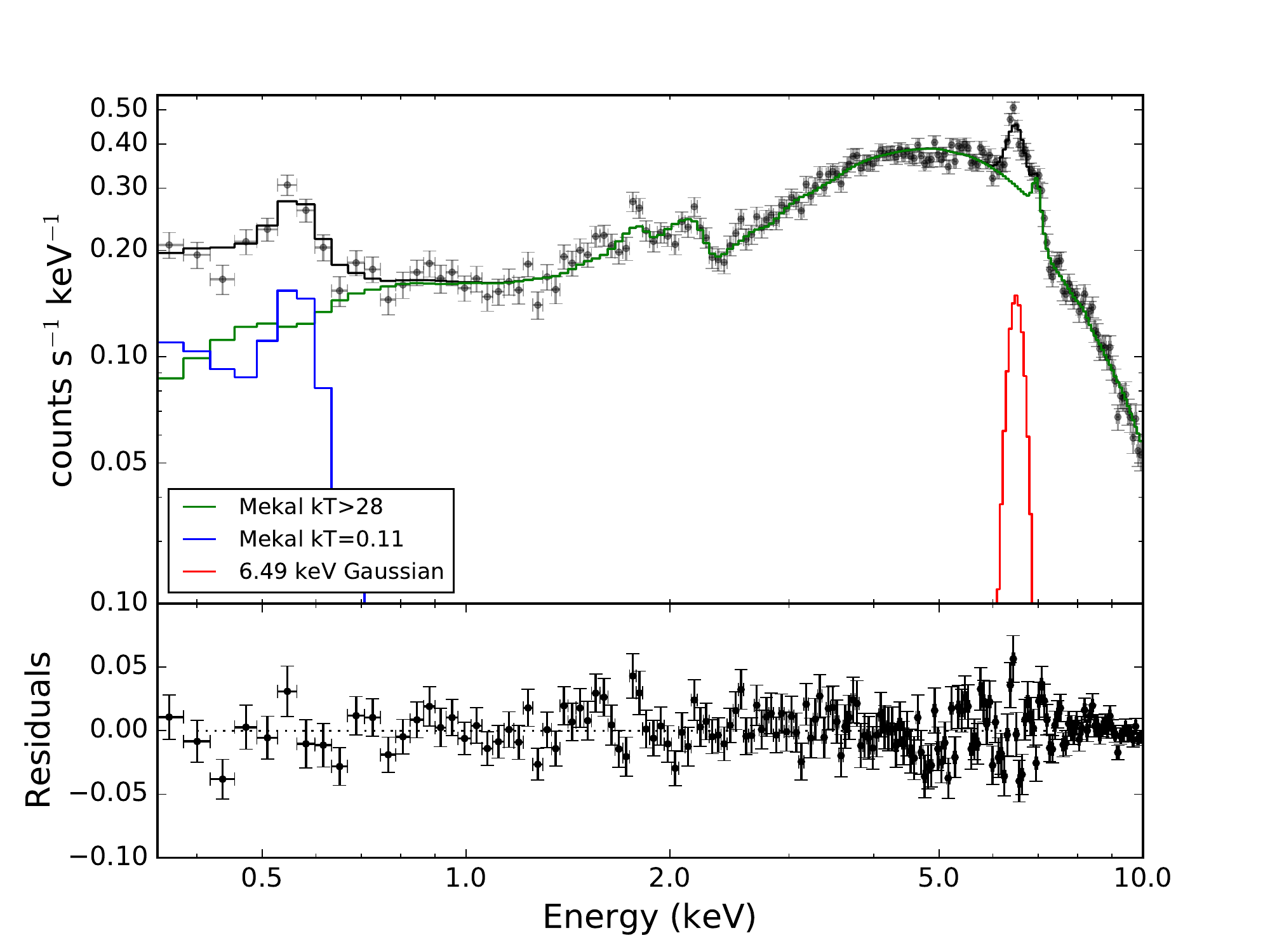}
    \caption{The X-ray model (black histogram) as described in the text for the EPIC-pn data, taken in 2016 when FO Aqr was in the recovery state. The individual components of the model are also shown.}
    \label{fig:model_xmm}
\end{figure}

Again, the best fit failed to constrain many of the parameters. A simplified version of the model used by \cite{Evans2004} was next used: a two {\sc mekal} model with a interstellar absorber and two circumstellar absorbers ({\sc Tbabs*(Tbabs*Partcov)*(Tbabs*Partcov)*(\\Xsmekal+Xsmekal)}). While the initial fit was good, there were large residuals around the 6.4 keV Fe line. A {\sc Gaussian} component was added to the model to account for these residuals, as was also done by \cite{Evans2004}. This model fit the three spectra with a $\chi^{2}_{R}=1.33$, and can be seen for the EPIC-pn spectrum in Fig.~\ref{fig:model_xmm}. The best fit parameters are given in the last column of Table~\ref{tab:model_data}. The unabsorbed flux in the 0.3-10 keV band for the recovery state model was $(7.2\pm0.6)\times10^{-11}$ erg cm$^{-2}$ s$^{-1}$.

\section{Discussion}

\subsection{Long Term Recovery}
\begin{figure}
	\includegraphics[width=\columnwidth]{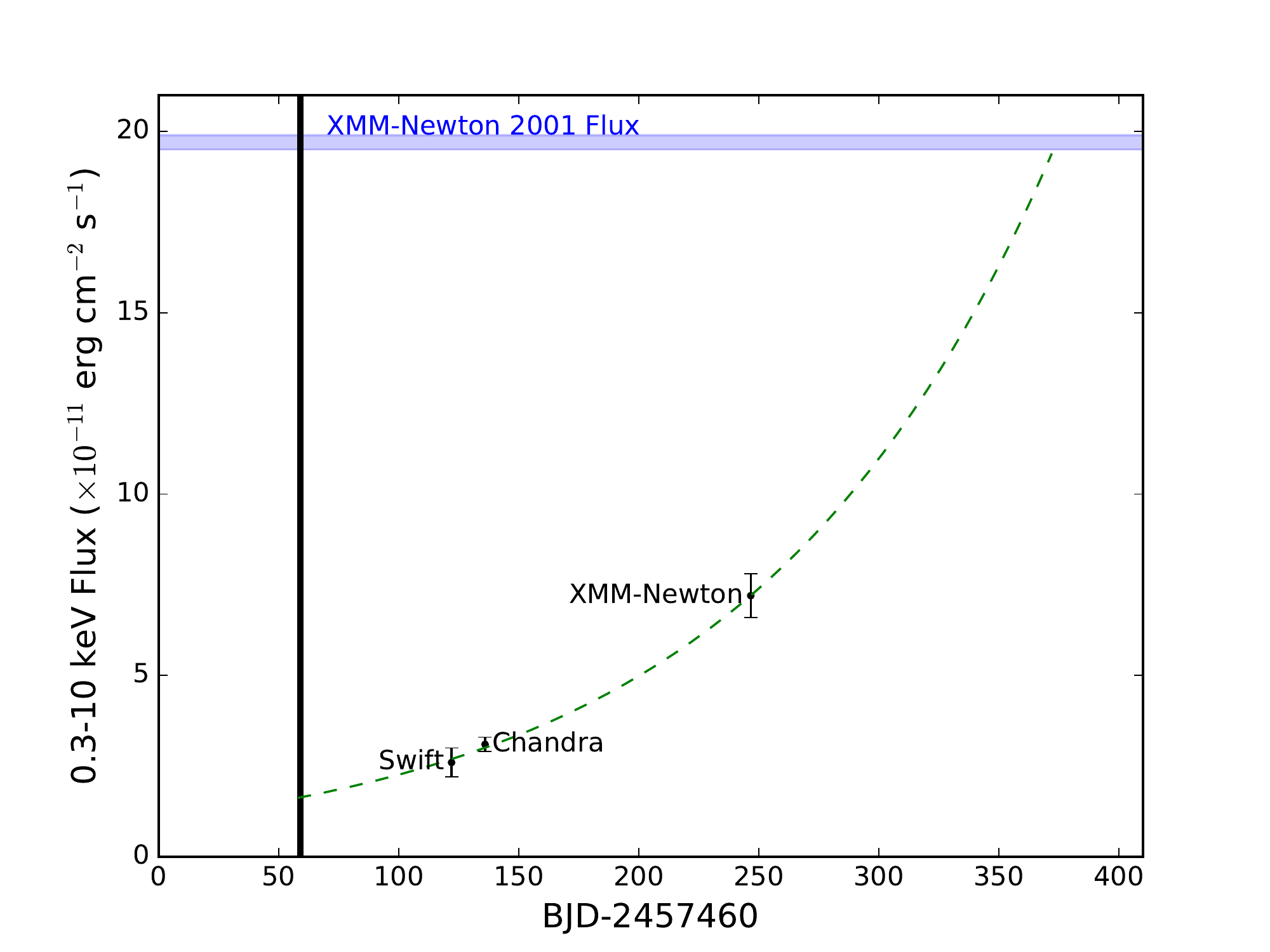}
    \caption{The long term X-ray recovery of FO Aqr. The blue shaded region shows the unabsorbed X-ray flux from the modelling of the 2001 \textit{XMM-Newton} data, when FO Aqr was in its high state. The black line shows the approximate time when FO Aqr was at its faintest. The green-dashed line shows the exponential fit from \citet{littlefield2016c} rescaled to fit the X-ray data, and is extended to show what the X-ray flux of FO Aqr should have been during faintest light, and when the X-ray flux should return to its normal high state.}
    \label{fig:longterm_recovery}
\end{figure}

Fig.~\ref{fig:longterm_recovery} shows the unabsorbed 0.3-10 keV X-ray flux from the low state observations taken by \textit{Swift} and \textit{Chandra} and from the recovery state observations taken by \textit{XMM-Newton}, alongside the unabsorbed flux of $(19.7\pm0.2)\times10^{-11}$ erg cm$^{-2}$ s$^{-1}$ from the 2001 \textit{XMM-Newton} observations. It is clear from this plot that by November 2016, FO Aqr had still not fully recovered from the low state. This is unsurprising as \cite{littlefield2016c} estimated the e-folding time (which is the time taken for the flux to change by a factor of e) for the optical light in FO Aqr to be 115$\pm$7 days. Based on a scaled fit of the same model used by \cite{littlefield2016c} to the X-ray fluxes shown in Fig.~\ref{fig:longterm_recovery}, we find that the e-folding time for the X-ray recovery is 124$\pm10$ days, which is in agreement with the optical recovery time from \cite{littlefield2016c}. This fit also suggests that if FO Aqr continues recovery uninterrupted, then it should return to its typical high state flux by March 2017.

Based on our spectral modelling and assuming the unabsorbed 0.3-10 keV X-ray flux is proportional to the mass accretion rate, the change in X-ray flux between the high state and the low state suggests that the mass accretion rate in the system decreased by a factor of $~ 7$ at the start of the low state.

\subsection{The softened X-ray spectrum}
The power spectrum of the \textit{Chandra} X-ray light curves and the energy spectrum of the \textit{Chandra} and 2016 \textit{Swift} data show a significant change when compared to the high state \textit{XMM-Newton} and \textit{Swift} data, and also when compared to the recovery state \textit{XMM-Newton} data. The power spectrum of the \textit{Chandra} data shows a near-equal distribution in power between the soft and hard X-rays at the spin frequency of the WD. There is also evidence of a strong signal at the beat period, which was not detected in the 2001 \textit{XMM-Newton} observations. Additionally, the $2\omega - 2\Omega$ and $2\omega - \Omega$ peaks show strong power in the hard X-rays and the $2\omega$ peak shows strong power in the soft X-rays in 2016, compared to the 2001 data, when only the $2\omega$ was detected, and only in the hard X-rays.

The interpretation of the change in the power spectrum is not an easy task. Based on the power spectra of optical data taken in the low state, \cite{littlefield2016c} suggested that FO Aqr had possibly transitioned into a ``disc-overflow'' accretion geometry. Following the analysis in both \cite{Wynn1992} and \cite{Ferrario1999}, the strong power at $\omega$ in 2001 is indicative of disc-fed accretion, while the strong power at $2\omega-\Omega$ and the power visible at $\omega-\Omega$ and $\Omega$ in the 2016 \textit{Chandra} supports the claim by \cite{littlefield2016c} that accretion had switched primarily to the stream-fed geometry.

X-rays with an energy $>5$ keV in IPs are thought to be generated at the top of the accretion column and are rarely absorbed by circumstellar material, while lower energy X-rays which come from the accretion flow close to the surface of the WD are subject to absorption through the accretion flow itself (this is called the self-absorption accretion curtain model; \citealt{1988MNRAS.231..549R}). The drop in hard X-ray flux visible in the 2016 \textit{Swift} spectrum when compared to the 2006 spectrum suggests less hard X-rays were being produced, leading to the conclusion that the mass accretion rate had decreased. The increase in the soft X-ray flux suggests there was less material in the curtains and whatever remains of the accretion disc to absorb soft X-rays. This is supported by the modelling of the various datasets, which shows a much lower circumstellar absorption density ($N_{H} = 1.3^{+0.6}_{-0.3}\times10^{22}$ cm$^{-2}$) in 2016 than in 2006 ($N_{H} = 7.2^{+1.2}_{-0.9}\times10^{22}$ cm$^{-2}$; \textit{Swift}) and in 2001 ($19.0^{+1.2}_{-0.9}\times10^{22}$ cm$^{-2}$; \textit{XMM-Newton}).

\subsection{The Recovery State}

Comparing the power spectrum of the \textit{XMM-Newton} data taken during the recovery state with the low state \textit{Chandra} data and the high state \textit{XMM-Newton} data suggests that FO Aqr had not fully returned to its typical high state by November 2016. The signal at $2\omega - \Omega$ visible in the \textit{Chandra} data during the low state had disappeared by the time the \textit{XMM-Newton} data were taken, and in the hard X-rays, the power spectrum of the recovery state resembled the power spectrum of the high state data from 2001. However, the power spectrum of the soft X-rays from the recovery state still showed significant power at the $\omega$ peak, which was not visible in the high state.

The implications of the recovery state power spectrum on the accretion geometry are important. The orbital modulation, $\Omega$, was detectable in both the hard and soft X-ray power spectra. However, the power of the orbital signal was much lower than during the high state observations in 2001. The weaker orbital modulation in the recovery state data is consistent with the photoelectric absorption model of \cite{2005A&A...439..213P} as, due to the lower accretion rate, this impact region should be less dense, leading to less photoelectric absorption, which would weaken the orbital modulation. 

The loss of power at the $2\omega-\Omega$ and $\omega-\Omega$ peaks suggest that FO Aqr had switched back to a disc-fed accretion system when the 2016 \textit{XMM-Newton} observations were taken.  The signal at $\omega$ in the soft X-ray power spectrum also implies that the system had still not fully recovered. A comparison of the soft X-ray light curves from 2001 (Fig.~\ref{fig:xmm_lc_01}) and 2016 (Fig.~\ref{fig:xmm_lc_16}) shows that, during the high state, no soft X-ray pulses are visible, but in the recovery state, an occasional soft X-ray pulse is detected. However, in the recovery state, strong soft X-ray pulses were never detected around orbital phase $\phi=0$, suggesting an orbital phase dependence of the material which is absorbing the soft X-rays.

\begin{figure}
	\includegraphics[width=\columnwidth]{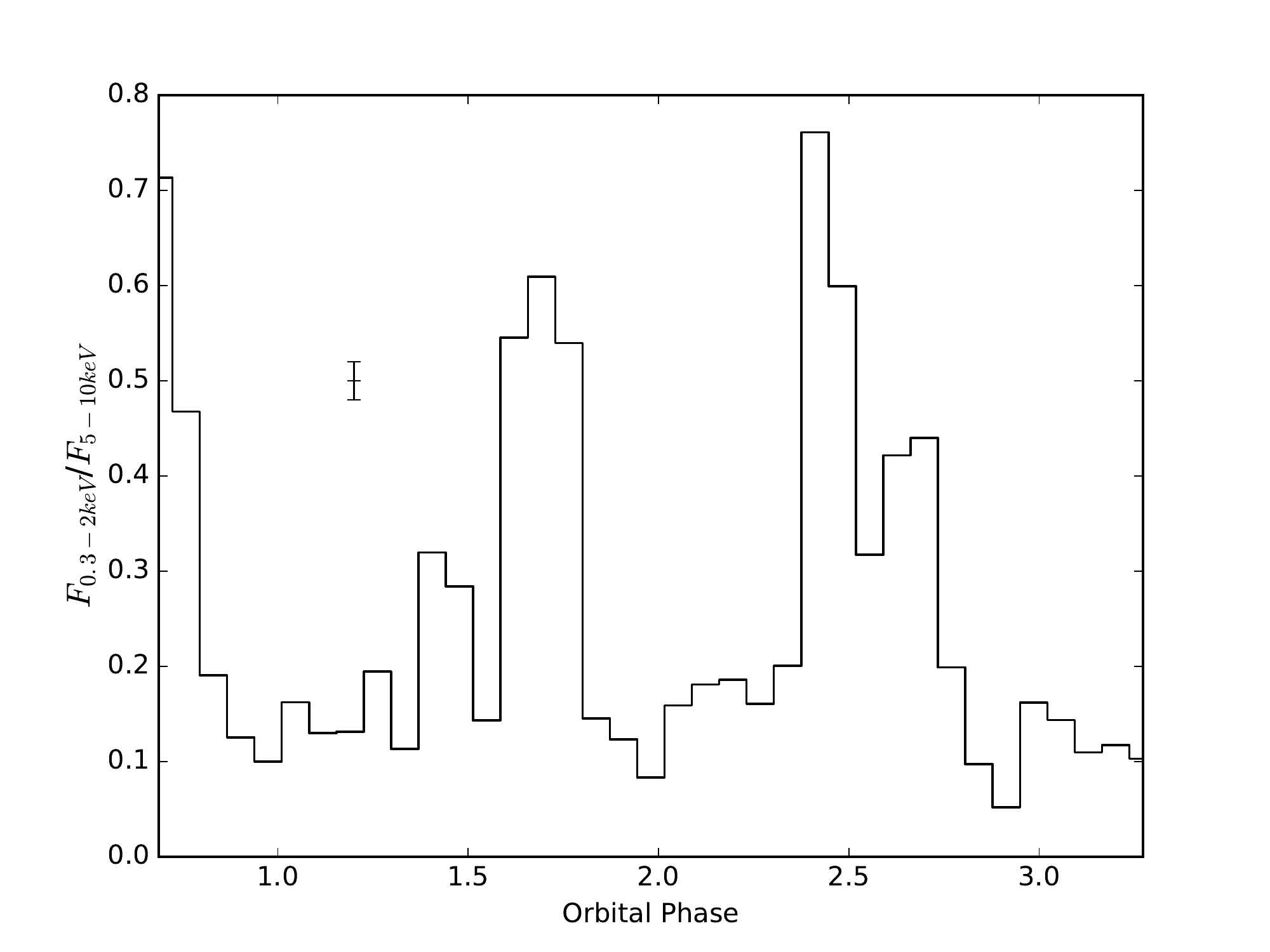}
    \caption{The softness ratio, defined as the flux in the 0.3-2 keV band divided by the flux in the 5-10 keV band, versus the orbital phase for the \textit{XMM-Newton} observations taken during the recovery state in 2016. The softness ratio displays a minimum around orbital phase 0, and show maxima when soft X-ray pulses were visible in the 0.3-2 keV light curve. The data have not been folded on orbital phase.}
    \label{fig:soft_hard_ratio}
\end{figure}

To better see this, light curves were extracted from the EPIC-pn instrument in the 0.3-2.0 keV and the 5.0-10.0 keV bands, and the ratio of these bands calculated for the duration of the \textit{XMM-Newton} observation. Fig.~\ref{fig:soft_hard_ratio} shows this softness ratio versus the unfolded orbital period. The light curve was binned to 1254.3343 s, the spin period of the WD, to remove effects of the spin pulse on this plot. 

From Fig.~\ref{fig:soft_hard_ratio}, it is clear that the softness ratio in FO Aqr varies over the orbital period, but does not vary by the same amount during each orbit. For example, during the first full orbit observed by \textit{XMM-Newton}, the softness ratio peaks with a value of 0.60$\pm0.02$ at orbital phase 0.7, while in the next full orbit, the softness ratio peaks with a value of 0.76$\pm$0.3 at orbital phase 0.4. However, for all 3 orbits observed, the softness ratio reaches a minimum around or just before orbital phase 0.0.

\begin{figure}
		\includegraphics[width=\columnwidth]{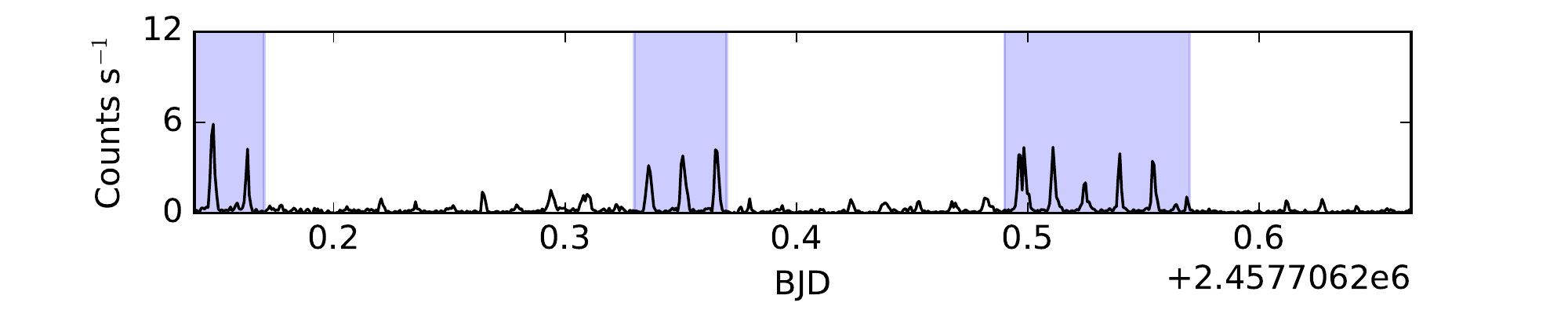}
	\includegraphics[width=\columnwidth]{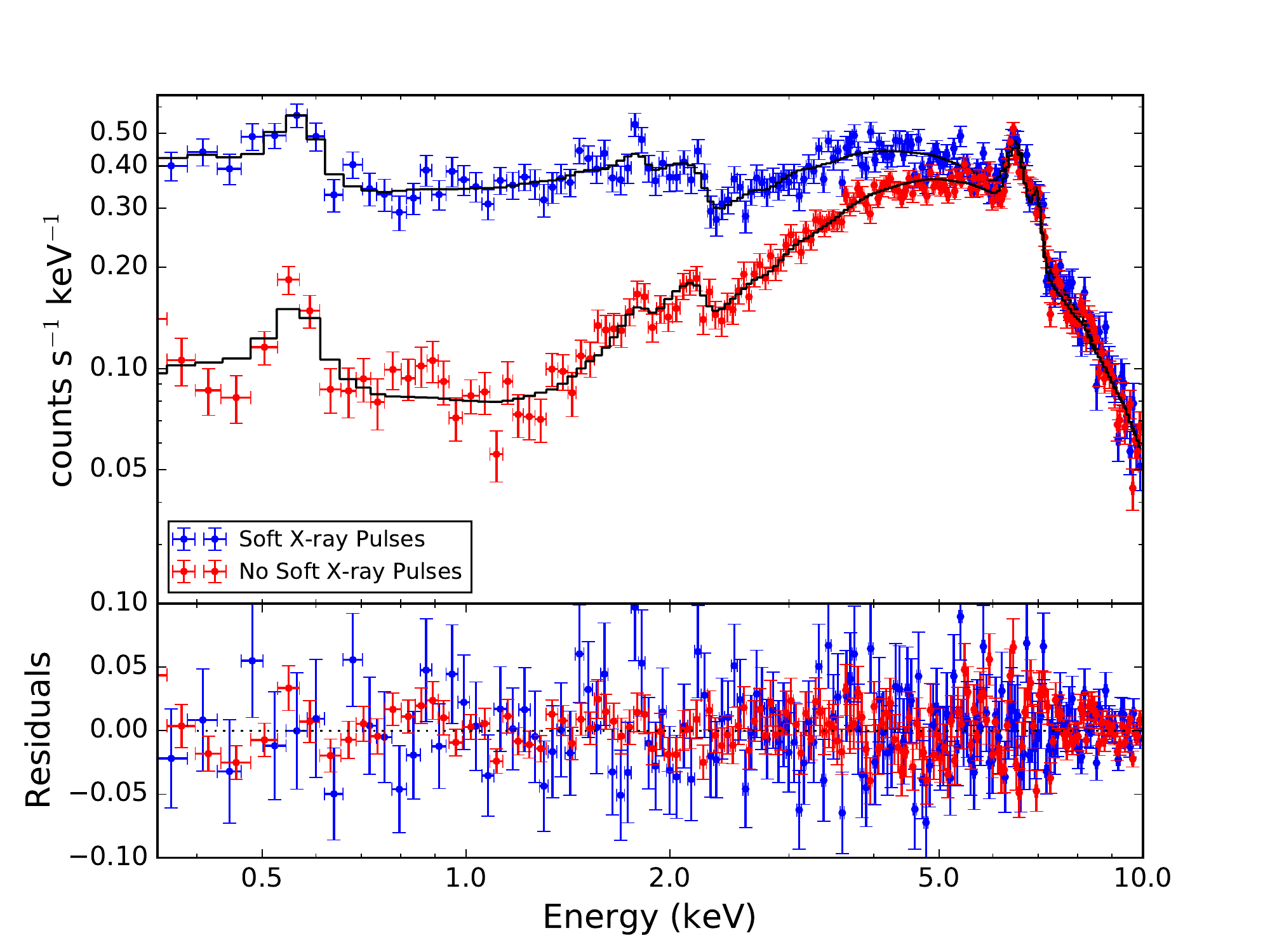}
    \caption{The top panel shows the 0.3-2 keV light curve from \textit{XMM-Newton}. The middle panel shows the spectrum extracted from the highlighted times in the top panel (upper curve, blue) and the spectrum extracted when all other events are used (bottom curve, red). The blue spectrum has a significantly softer component than the red spectrum, which is characterised by lower partial-covering fraction in the partial absorbers (see Table \ref{tab:model_soft_xrays}).}
    \label{fig:soft_pulse_model_xmm}
\end{figure}

To better quantify the changes in the spectrum between periods when soft X-ray pulses are and are not visible, the EPIC-pn, -MOS1 and -MOS2 data were split up. A spectrum was constructed out of events which arrived when soft X-ray pulses were visible in the recovery state light curve (see the highlighted regions in the top panel of Fig.~\ref{fig:soft_pulse_model_xmm}). Another spectrum was constructed out of events which arrived at all other times (during the good time interval). Both of the EPIC-pn spectra are visible in the middle panel of Fig.~\ref{fig:soft_pulse_model_xmm}, and show that\ as expected, the spectrum of FO Aqr when these soft X-ray pulses are visible has a significantly stronger soft component compared with when there are no soft X-ray pulses. The same model that was fit to the spectrum in Section~\ref{sec:XMM_Model} was fit to the -pn, -MOS1 and -MOS2 spectra for these separate time intervals. The results of the fitting can be seen in Table~\ref{tab:model_soft_xrays}.

\begin{table*}
	\centering
	\caption{The parameters for modelling the two spectra shown in Fig.~\ref{fig:soft_pulse_model_xmm}. The two models have nearly identical components, with the exception of the covering fractions for the partial absorbers. The change in the partial absorbers is thought to be the reason for the appearance of the occasional soft X-ray pulse in the 0.3-2 keV band of the recovery state data.}
	\begin{tabular}{r  c c c c}
		\hline
		Component 						& Parameter							& Soft pulses   	                    & No Soft pulses\\
		\hline\hline
		{\sc tbabs (interstellar)}		& $N_{H}$ (10$^{22}$ cm$^{-2}$)		& 0.10$\pm$0.07                         & 0.07$^{+0.09}_{-0.06}$ \\
		\\
		{\sc tbabs (circumstellar 1)}	& $N_{H}$ (10$^{22}$ cm$^{-2}$)		& 13$^{+6.0}_{-4.0}$                    & 15$^{+3.0}_{-2.0}$\\ 
		\\
		{\sc partcov 1}					& cvf								& 0.65$^{+0.09}_{-0.11}$                & 0.74$\pm$0.05\\
		\\
		{\sc tbabs (circumstellar 2)}	& $N_{H}$ (10$^{22}$ cm$^{-2}$)		& 2.9$\pm$1.0                           & 3.9$^{+0.7}_{-0.6}$\\ 
		\\
		{\sc partcov 2}					& cvf								& 0.81$^{+0.04}_{-0.07}$                & 0.95$\pm$0.01\\
		\\
		\multirow{3}{*}{\sc Mekal}		& kT (keV)							& $>$20                                 & $>$ 29\\
		                                & Abundance                         & 0.3$^{+0.5}_{-0.25}$                   & 0.5$^{+0.4}_{-0.3}$\\
										& norm		        	            & 0.036$^{+0.005}_{-0.003}$             & 0.037$\pm$0.002\\
		\\
		\multirow{3}{*}{\sc Mekal}		& kT (keV)							& $<$0.13                               & 0.12$^{+0.05}_{-0.03}$\\
		                                & Abundance                         & 0.3$^{+0.5}_{-0.25}$                   & 0.5$^{+0.4}_{-0.3}$\\
										& norm		        	            & 0.05$^{+0.2}_{-0.04}$               & 0.02$^{+0.05}_{-0.01}$ \\
		\\
		\multirow{3}{*}{\sc Gaussian}	& Centre							& 6.49$^{+0.07}_{-0.05}$                & 6.05$^{+0.05}_{-0.04}$\\
		                                & Sigma                             & 0.17$^{+0.10}_{-0.07}$                & 0.18$^{+0.08}_{-0.05}$ \\
										& norm ($\times10^{-4}$)            & 1.3$^{+0.5}_{-0.4}$                   & 1.3$\pm$0.3  \\
		
										& $\chi^{2}_{R}$					& 1.23                                  & 1.38\\
		\hline
	\end{tabular}
	\label{tab:model_soft_xrays}
\end{table*}

The two models in Table~\ref{tab:model_soft_xrays} are nearly identical - the interstellar absorber, {\sc Mekal} values and {\sc Gaussian} values all lie within 3$\sigma$ of each other in both fits. However, there is a significant difference in the partial covering fraction of the second circumstellar absorber between the two models. When the soft X-ray pulses are present, the partial covering fraction is 0.81$^{+0.04}_{-0.07}$, while when no soft X-ray pulses are present, this covering fraction is 0.95$\pm$0.01 (the errors here are quoted at the 3$\sigma$ level). This suggests that for part of the orbit, the region which is producing the soft X-ray component is obscured by an optically thick region.

\subsection{X-ray and Optical Pulse}
\subsubsection{Low State}
\begin{figure}
	\includegraphics[width=\columnwidth]{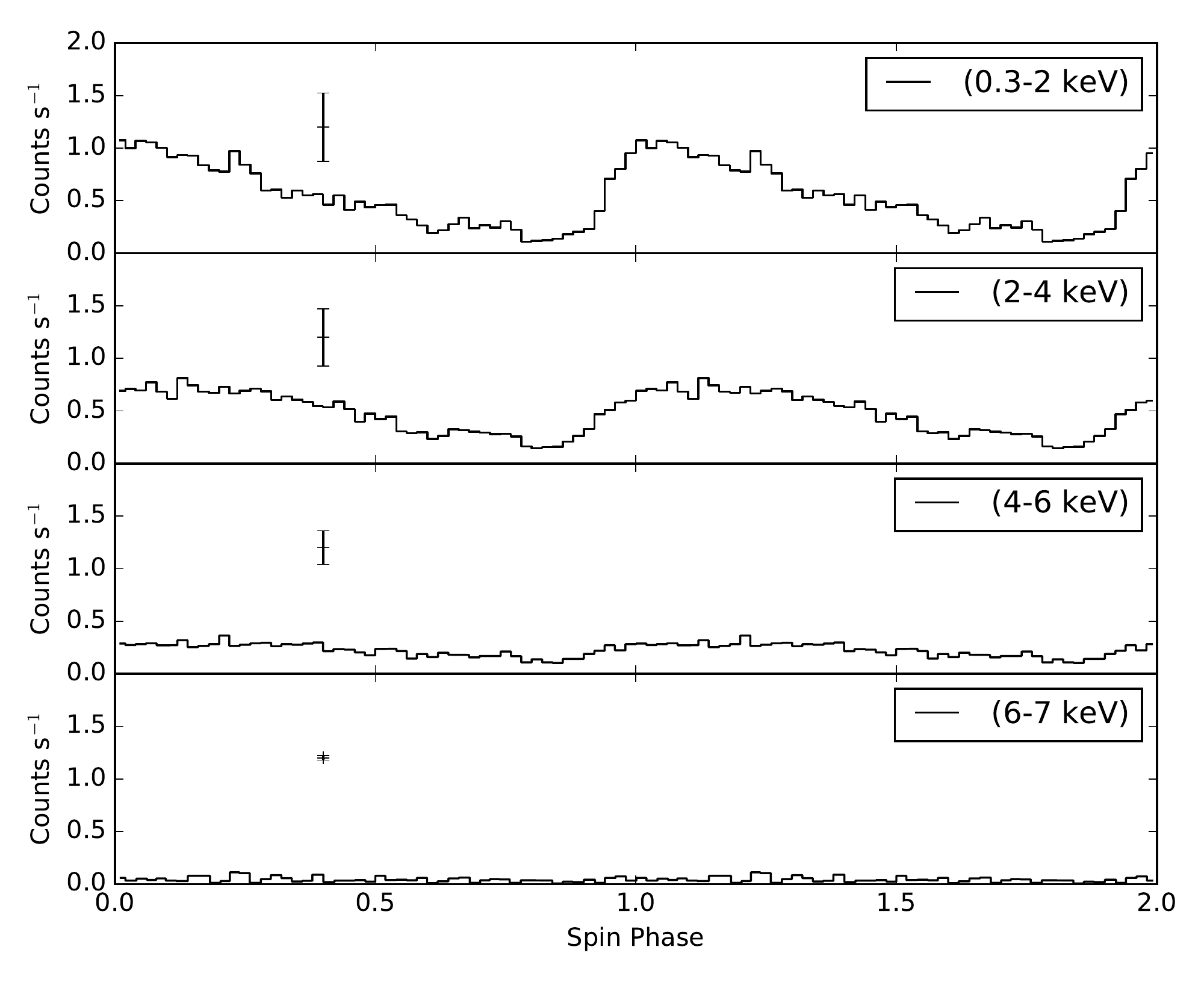}
    \caption{Various X-ray light curves taken by \textit{Chandra} in July 2016. There is a clear X-ray pulse visible in the 0.3-2 and 2-4 keV bands. The rise time of the pulse is faster than the decay time, suggesting the X-ray producing region is extended in the azimuth direction close to the WD.}
    \label{fig:split_x_ray_lcs_chandra}
\end{figure}

Fig.~\ref{fig:split_x_ray_lcs_chandra} shows the X-ray light curve for the 0.3-2 keV, 2-4 keV, 4-6 keV and 6-7 keV energy bands phased using Equation \ref{eqn:SpinEph}. The X-ray pulse is most obvious in the 0.3-2 keV band, and has a decaying amplitude at increasing energies. The shape of the pulse is also quite peculiar - the pulse has a very fast, sharp rise starting around spin phase 0.92. and ending around phase 1.00, while the decay is gradual and slow, lasting from spin phase 1.00 until the rise starts again. This suggests the region producing the soft X-rays comes into view very quickly, and then is gradually eclipsed by material of increasing optical depth.

\subsubsection{Recovery State}

The arrival times of the optical and X-ray pulses seen in the 2016 \textit{XMM-Newton} data are similar, suggesting the origin of the optical and X-ray pulses are the very close together. \cite{Hellier1990} calculated the \textit{V/R} ratio for the optical emission lines (which is simply the ratio of the equivalent widths on either side of the line rest wavelength) and found that at spin maximum, the \textit{V/R} ratio also has a maximum which suggests that whatever material is producing the emission must be streaming towards the observer at this phase. To explain the similar arrival of the pulses in the optical and X-rays, along with the observed \textit{V/R} for the emission lines at the phase maximum, \cite{Hellier1990} proposed that the optical and X-ray modulations in FO Aqr are caused by ``self-absorption'' by the optically thick accretion curtain. In this scenario, the maxima of the optical and X-ray pulse are seen when the magnetic pole closest in orientation to us is pointed away from the Earth, allowing us to see a maximum area of the part of the accretion curtain which is producing the optical and soft X-rays, while the minima of the pulse is seen when this pole is pointed towards us, but through the accretion curtain which is now absorbing a majority of the produced radiation. This scenario has also been employed to explain the simultaneous arrival of the optical and X-ray pulses in EX Hya (\citealt{1987MNRAS.228..463H}; \citealt{1988MNRAS.231..549R}; \citealt{1991MNRAS.249..417R}). For the rest of this paper, we refer to this pole as the ``top" magnetic pole, while the magnetic pole which is orientated furthest from Earth is the ``bottom'' pole. For a diagram of this geometry, see Fig 2 of \cite{Wynn1992}.

\begin{figure}
	\includegraphics[width=\columnwidth]{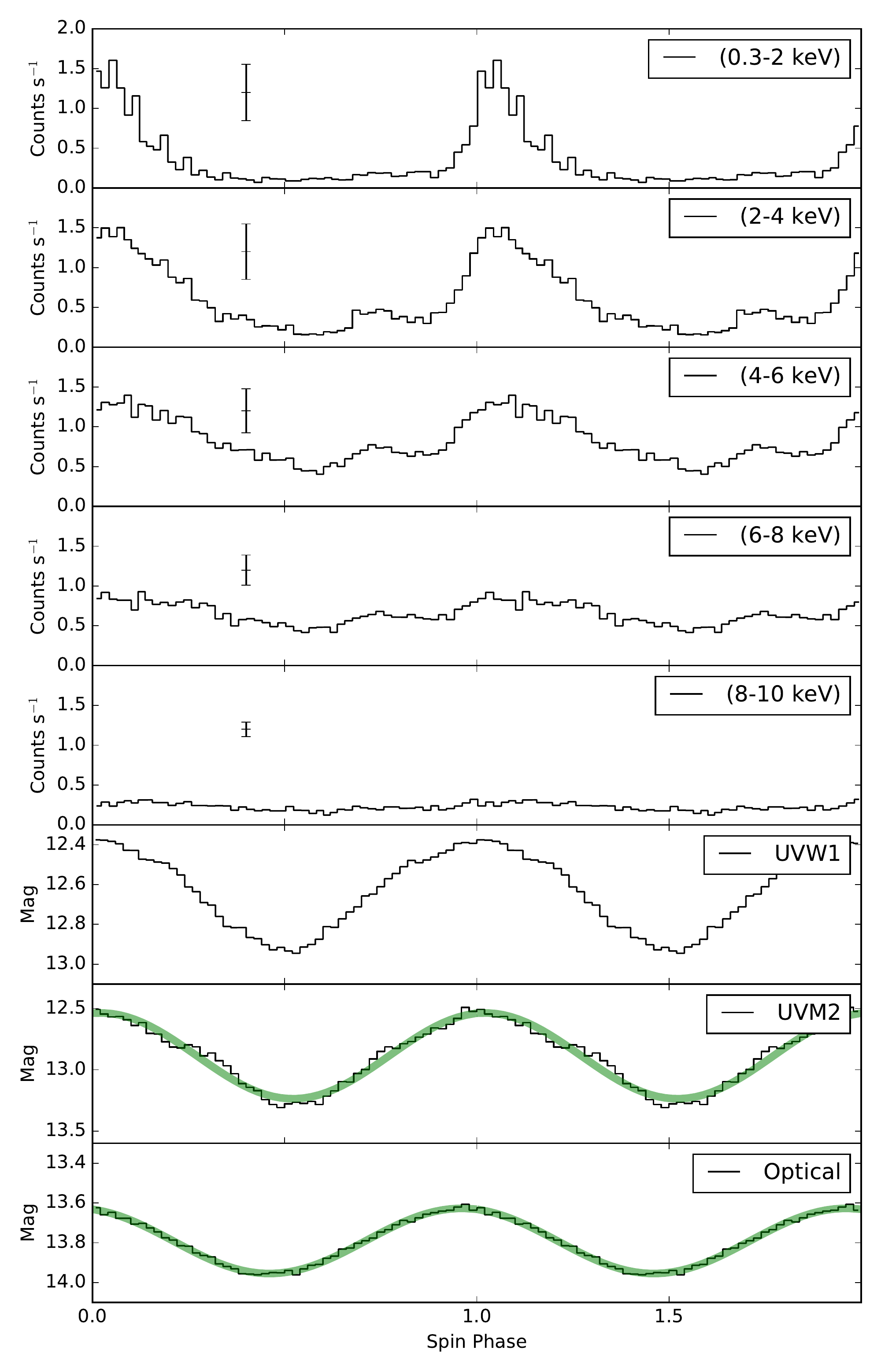}
    \caption{Various X-ray light curves taken by \textit{XMM-Newton} in November 2016. The X-rays with energies 2 keV$<$E$<$8 keV are pulsed, while hard X-rays with energy $>$ 8 keV are not. The pulsed X-rays are in phase with the optical pulses, suggesting the ``self-absorption'' model of accretion curtains is leading to the pulsed emission in FO Aqr. The pulses in the 0.3-2 keV band are not always present, but are orbital phase dependant, as shown in Fig.~\ref{fig:xmm_lc_16}. The green line in the bottom two plots shows the sine curve fit to the UV and optical pulses.}
    \label{fig:split_x_ray_lcs}
\end{figure}

To check if this scenario still applies to FO Aqr, the X-ray light curve was split up into several energy bands (0.3-2 keV, 2-4 keV, 4-6 keV, 6-8 keV and 8-10 keV) and phased on the spin period. The resulting light curves are shown in Fig.~\ref{fig:split_x_ray_lcs}, and show that the X-ray pulsations are strongest in the 2-4 keV band, and undetectable in the 8-10 keV band, confirming that the ``self-absorption'' model invoked by \cite{Hellier1990} applies to the recovery state. 

The constant flux from the 8-10 keV band suggests that the accretion region which is producing the hard X-rays around the top magnetic pole is always visible from Earth, or that when this location is hidden from Earth, the accretion region around the bottom magnetic pole becomes visible, which helps maintain the hard X-ray flux.

The shape of the X-ray pulse in the recovery state is very different to the pulse seen in the low state. The rise of the X-ray pulse is slower than in the low state, starting at spin phase 0.92 and reaching a maximum at 0.05, but the decay is much faster, reaching a constant flux level by spin phase 0.3. This suggests the material absorbing the X-ray pulses is more symmetric in the recovery state than in the low state.

\begin{figure}
	\includegraphics[width=\columnwidth]{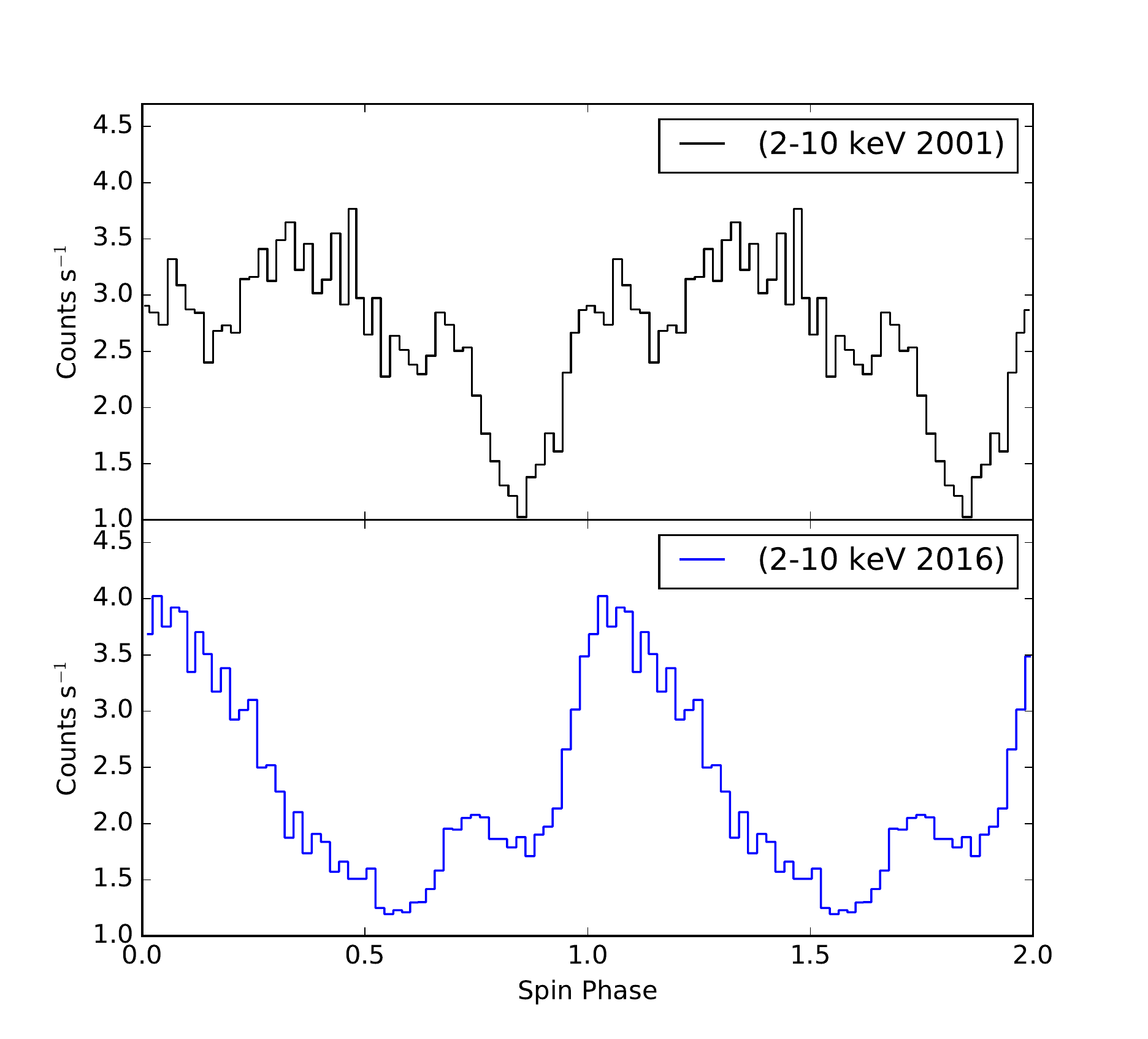}
    \caption{The spin-phased 2-10 keV X-ray pulses observed during the high (black, top) and recovery (blue, bottom) observations. The X-ray pulses show three distinct features (a minimum, maximum and notch) in both light curves, but with the recovery features phase shifted relative to the high state features.}
    \label{fig:comp_01_10}
\end{figure}

Fig.~\ref{fig:comp_01_10} shows the spin-phased light curves from the recovery data and the high state data. The high state X-ray observations from \cite{Evans2004} were phased using their spin ephemeris, since the spin ephemeris here would have had a large error associated with it if propagated back to 2001. It is important to note that both X-ray light curves are phased based off of ephemerides derived from the optical pulsations. 

The X-ray pulse in the recovery state seems to share many of the features seen in the high state pulse, except they are reduced and phase shifted. The 3 main features identified by \cite{Evans2004} in the high state X-ray pulse were the minimum at spin phase $\phi \sim 0.67-0.87$, the `notch' at spin phase $\phi \sim 0.99-1.11$ and the maximum at spin phase $\phi \sim 0.12-0.36$. The minimum and maximum are both apparent in the recovery state spin pulse, but are at spin phases $\phi \sim 0.51-0.68$ and $\phi \sim 0.0-0.14$ respectively. This corresponds to a phase shift of 0.17 between the high and recovery state pulse. The `notch' is also present in the recovery spin pulse. However, it is much weaker, which is most likely related to the lower count rate right before the notch in the recovery state at spin phase $\phi \sim 0.67-0.78$, and also phase shifted to $\phi \sim 0.80-0.90$. The apparent phase shift in the X-ray pulse is striking, since the UV pulses observed during both the high and recovery state data arrive at  $\phi \sim 0.0$. This means that during the high state, the phase difference between the arrival of the maximum of the UV pulse and the maximum of the X-ray pulse was $\sim 0.12-0.36$, while the phase difference during the recovery state was much less, closer to $\sim 0.0-0.14$.

\cite{Evans2004} suggested that the magnetic field lines in FO Aqr may be twisted such that the infalling material is trailing the accreting magnetic pole. In their interpretation of the X-ray pulse, the `notch' is caused by the brief disappearance of part of the X-ray emitting region, and that the minimum is caused by absorption of the X-ray flux coming from the lower magnetic pole by the upper curtain. For a symmetric magnetic field, it would then be expected that the phase difference between the `notch' and the minimum is $\sim 0.5$. However, \cite{Evans2004} noted that the spin phase difference between the `notch' and the minimum was closer to $\sim0.77$ in the high state, which was their evidence for the twisted magnetic field lines. During the time of the high state observations, FO Aqr was spinning up (the spin period was decreasing; \citealt{Williams2003}). Another object of interest is PQ Gem, which is an IP that is spinning down (spin period is increasing; \citealt{1997MNRAS.285..493M}; \citealt{2006MNRAS.369.1229E}) and whose magnetic field lines are also twisted, but in the opposite direction such that the infalling material is leading the accreting magnetic pole \citep{2006MNRAS.369.1229E}.

\cite{2014EPJWC..6407001H} suggested that the trailing and leading of infalling material may be related to the spin up and spin down of these systems, based on FO Aqr and PQ Gem. While this hypothesis has not been fully tested, it does provide an explanation for the apparent phase difference in the X-ray observations of FO Aqr between the high and recovery state. As \cite{Kennedy2016} showed, FO Aqr is no longer spinning up, but has transitioned to spinning down. It may be that the observed X-ray phase change is related to the flip in the sign of $\dot{P}$, and is indicative of a change in the twist in the magnetic field lines. We do note that the phase separation of $\phi \sim 0.8$ between the notch at $\phi \sim 0.80-0.90$ and the minimum at $\phi \sim 0.51-0.68$, which is what \cite{Evans2004} used to suggest the magnetic field lines were twisted, in the recovery spin pulse is the same as the phase separation in the high state. This suggests the twist in the magnetic field may not have changed at all. However, since the spin pulse in the recovery state has a slightly different shape (the high state pulse was significantly broader, as shown in Fig.~\ref{fig:comp_01_10}) and the notch in particular was less obvious our ability to measure this phase difference accurately is limited. One thing for certain is that further X-ray observations of FO Aqr must be carried out to identify the long term nature of this apparent X-ray phase shift.

The UV and optical pulses shown in the bottom two panels of Fig.~\ref{fig:split_x_ray_lcs} are out of phase with each other. This phase lag was measured to be 0.062$\pm$0.001 in spin phase units by modelling the optical and UV pulses as sine functions (these models are also shown in Fig.~\ref{fig:split_x_ray_lcs}). Since the optical and UV pulse comes from the pre-shock accretion flow, it is likely that this lag is associated with a twist in the accretion curtains, as proposed by \cite{Evans2004}.

\section{Conclusions}
We have presented the first-ever X-ray observations of an IP during a low state, along with X-ray observations during the subsequent recovery. In the low state, the X-ray spectrum suggests the accretion rate within FO Aqr dropped significantly, leading to a lower production of hard X-rays. Additionally, the lower accretion rate led to a lower column density within the accretion structures increasing the observed soft X-ray flux. The power spectrum of the X-ray data suggests that mass transfer via an accretion stream was the dominant accretion mode in the low state.

The recovery spectra suggest that, as of November 2016, FO Aqr still had not fully returned to its typical high state. While the power spectrum of the hard X-rays in the recovery state resembled those obtained in the high state, implying the accretion geometry had returned to a disc-fed scenario, the 0.3-2 keV light curve showed occasional pulses, which are absent in the archive X-ray data from 2001. Detailed analysis of the X-ray spectrum suggests that the densities within the accretion disc and curtains had increased since the low state observations in July. The appearance of the occasional soft X-ray pulse between orbital phases 0.3-0.8 suggests there are still areas of low density within the accretion structure, allowing occasional glimpses into the soft X-ray emitting regions of the system.

When compared with the X-ray pulse observed in the high state, the recovery X-ray pulse shows many similar features, albeit with a phase shift. This suggests a possible change in the structure of the accretion flow which is generating the soft X-rays. However, whether this phase shift is now a permanent feature of FO Aqr, or was only visible in the recovery state remains to be seen.

\section*{Acknowledgements}

We thank Neil Gehrels for granting target of opportunity observations with the \textit{Swift} satellite. Sadly Dr. Gehrels passed away before this research was complete. We acknowledge his tremendous contribution to many fields of astrophysics. We would like to thank Belinda Wilkes, director of the \textit{Chandra} X-ray Center, for granting us Director Discretionary Time on the \textit{Chandra} X-ray Telescope and Norbert Schartel and the \textit{XMM-Newton} OTAC for granting us TOO observations of FO Aqr. We would also like to thank the AAVSO for support observations during the \textit{XMM-Newton} observations. In particluar, we would like to thank Teofilo Arranz, Michelle Dadighat, Emery Erdelyi, James Foster, Nathan Krumm, David Lane, Damien Lemay, Diego Rodriguez Perez, Geoffrey Stone and Brad Vietje. MRK, PC and PMG acknowledge financial support from the Naughton Foundation, Science Foundation Ireland and the UCC Strategic Research Fund. MRK and PMG acknowledge support for program number 13427 which was provided by NASA through a grant from the Space Telescope Science Institute, which is operated by the Association of Universities for Research in Astronomy, Inc., under NASA contract NAS5-26555. STSDAS and PyRAF are products of the Space Telescope Science Institute, which is operated by AURA for NASA. This material is based upon work supported financially by the National Research Foundation. Any opinions, findings and conclusions or recommendations expressed in this material are those of the author(s) and therefore the NRF does not accept any liability in regard thereto. This research made use of Astropy, a community-developed core Python package for Astronomy \citep{2013A&A...558A..33A}.




\bibliographystyle{mnras}
\bibliography{FOAQr2} 





\bsp	
\label{lastpage}
\end{document}